\documentclass[a4paper,11pt]{article}
\pdfoutput=1

\usepackage{jheparxiv} 
\usepackage[T1]{fontenc}
\usepackage[utf8]{inputenc}
\usepackage{graphicx}
\usepackage{amsmath}
\usepackage{amssymb}
\usepackage{mathtools}
\usepackage{dsfont}
\usepackage{latexsym}
\usepackage{mathrsfs}
\usepackage{braket}		
\usepackage{hyperref}
\usepackage[dvipsnames]{xcolor}
\usepackage{twistor}

\newcommand{\eps}{\epsilon}
\newcommand{\veps}{\varepsilon}

\newcommand{\pa}{\partial}

\newcommand{\hb}{\bar{h}}

\renewcommand{\bz}{\bar z}
\newcommand{\bh}{\bar h}
\newcommand{\rO}{\mathrm{O}}
\newcommand{\w}{\mathrm{w}}

\title{All-order celestial OPE from on-shell recursion}
\author[a]{Lecheng Ren,}
\author[b]{Anders Schreiber,}
\author[c]{Atul Sharma,}
\author[d]{Diandian Wang}
\affiliation[a]{Department of Physics, Brown University, \\Providence, RI 02912, USA \vspace{0.1cm}}
\emailAdd{lecheng\_ren@brown.edu}
\affiliation[b]{Max-Planck-Institut für Physik, Werner-Heisenberg-Institut,\\ 80805 München, Germany\vspace{0.1cm}}
\emailAdd{ohrbergs@mpp.mpg.de}
\affiliation[c]{Center for the Fundamental Laws of Nature \& Black Hole Initiative,\\Harvard University, Cambridge, MA 02138, USA \vspace{0.1cm}}
\emailAdd{atulsharma@fas.harvard.edu}
\affiliation[d]{Department of Physics, University of California,\\ Santa Barbara, CA 93106, USA}
\emailAdd{diandian@physics.ucsb.edu}

\abstract{We determine tree level, all-order celestial operator product expansions (OPEs) of gluons and gravitons in the maximally helicity violating (MHV) sector. We start by obtaining the all-order collinear expansions of MHV amplitudes using the inverse soft recursion relations that they satisfy. These collinear expansions are recast as celestial OPE expansions in bases of momentum as well as boost eigenstates. This shows that inverse soft recursion for MHV amplitudes is dual to OPE recursion in celestial conformal field theory.}

\begin{document}

\maketitle

\section{Introduction}

Holography in asymptotically flat spacetimes has grown into a vividly active field of research in recent years. An emerging candidate for such dualities is the idea of celestial holography \cite{Pasterski:2021raf}. According to its tenets, quantum gravity in zero cosmological constant spacetimes is proposed to be dual to \emph{celestial conformal field theories} (CCFT) living on lightcone cuts like the celestial sphere in Lorentzian signature \cite{Pasterski:2016qvg, Pasterski:2017kqt} or the celestial torus in split signature \cite{Atanasov:2021oyu}. Scattering amplitudes in the bulk are expected to equal correlators of local operators in such CCFTs. A top-down toy model realizing some of these expectations has also been put forward in \cite{Costello:2022jpg}.

A key observation made in \cite{Fan:2019emx} was that collinear limits of flat space scattering amplitudes map to operator product expansions (OPEs) in the dual CFT. In subsequent work \cite{Pate:2019lpp}, their observation was generalized to compute such \emph{celestial OPEs} in a wide variety of theories like Yang-Mills and Einstein gravity. They determined the singular terms in celestial OPEs by using universal collinear limits that hold for all gluon and graviton amplitudes. In this paper, we will be concerned with completing the program of computing these OPEs to all orders (singular as well as regular) in the collinear expansion at tree level.

At this stage in the subject, due to open issues regarding locality and associativity of the celestial OPE \cite{Mago:2021wje,Ren:2022sws,Ball:2022bgg}, this computation is only accessible in a specific scattering configuration called the maximally helicity violating (MHV) sector. This corresponds to scattering two negative helicity massless particles against multiple positive helicity massless particles. It is this configuration that we will restrict attention to. At tree level in the MHV sector, we will show that the all-order celestial OPE of gluons and gravitons can be very easily extracted purely from the inverse soft recursion relations that their amplitudes satisfy. 

This method bypasses a number of computational hurdles encountered in past work. In the original works \cite{Banerjee:2020kaa, Banerjee:2020zlg, Ebert:2020nqf, Banerjee:2020vnt} that started the study of subleading terms in celestial OPEs, the authors were restricted to working order-by-order in a non-systematic fashion. They only studied regular terms at the first and second subleading orders in the collinear limit, and were also restricted to low multiplicity amplitudes. This was because they worked with explicit expressions for the MHV amplitudes like the Parke-Taylor or Hodges' formulae. Mellin transforms of these formulae tend to be highly nontrivial generalized hypergeometric functions \cite{Schreiber:2017jsr}. We will show that it is incredibly more judicious to work directly with the recursion relations. In fact, the lesson one learns this way is that the inverse soft recursion for MHV amplitudes is \emph{equivalent} to the OPE recursion in celestial CFT. In this sense, on-shell recursion takes the guise of an emergent phenomenon. This point has already been hinted at by other works like \cite{Costello:2022wso} for CSW rules or \cite{Hu:2022bpa} for BCFW recursion.

The most powerful application of subleading terms in celestial OPEs has been to the discovery of infinite chiral symmetries. In a beautiful series of works \cite{Guevara:2019ypd, Guevara:2021abz, Strominger:2021lvk, Himwich:2021dau, Adamo:2021lrv}, it was gradually realized that celestial duals of gravity and gauge theory possess symmetries generated by the loop algebras of $\text{Ham}(\C^2)$ and $\text{Maps}(\C^2,\mathfrak{g})$ respectively, with $\mathfrak{g}$ being the gauge group's Lie algebra. Our results for the all-order celestial OPE will be adapted to their representation theory. Crucially, the only descendants that appear in our OPEs will be the descendants associated to these symmetries, along with antiholomorphic conformal descendants. There won't be any holomorphic conformal descendant in sight! The reason this does not shriek inconsistency is that holomorphic conformal descendants happen to be expressible in terms of such current algebra descendants due to null state relations \cite{Banerjee:2020vnt,Banerjee:2020zlg} (see also \cite{Banerjee:2021cly,Banerjee:2021dlm,Banerjee:2023zip}).

In the case of MHV gluon scattering, our all-order OPE has been previously obtained from a concrete vertex algebra realization in \cite{Adamo:2022wjo}. This specific realization was constructed by studying twistor string theory in the MHV sector. Twistor and ambitwistor strings are a class of topological string theories that give rise to remarkably compact worldsheet formulae for gluon and graviton tree amplitudes in all N$^k$MHV sectors (i.e., amplitudes involving $k+2$ negative helicity particles) \cite{Witten:2003nn,Roiban:2004yf,Skinner:2013xp,Mason:2013sva,Geyer:2014fka}. As such, the algebra of their vertex operators can be systematically matched to the celestial OPE algebra (see also \cite{Adamo:2021zpw, Bu:2021avc}). However, calculating celestial OPEs from such vertex operators becomes cumbersome rather quickly. For example, it has proven much trickier to repeat such a calculation for the all-order MHV graviton OPE. In contrast, our recursion relation based approach will display no such shortcomings.

In the rest of this work, we provide more details of this approach. In Section~\ref{sec:prel}, we review inverse soft recursion relations in the MHV sector. In Sections~\ref{sec:gluon} and \ref{sec:graviton}, we turn them into all-order collinear expansions for gluon and graviton scattering respectively. These are recast as all-order celestial OPEs both in momentum space and in the space of boost eigenstates. In Section~\ref{sec:disc}, we close with a discussion of possible applications and generalizations. 

Appendix \ref{app:sing} discusses the behavior of soft-hard correlators at infinity, which feeds into certain contour integral manipulations used to obtain correlators with descendant insertions. Appendix \ref{app:currentOPE} contains the computation of the all-order OPE of soft gluon and graviton currents, again valid within the MHV sector. Appendix \ref{app:computeOPE} provides further details of the Mellin transforms that convert momentum eigenstate celestial OPEs to those of boost eigenstates.

\section{Inverse soft recursion}
\label{sec:prel}

In this section, we review the inverse soft recursion relations for MHV amplitudes. Brief definitions of concepts of celestial holography will also be provided as and when required in later sections. Many reviews of celestial holography are by now available in the literature, see for instance \cite{Pasterski:2021rjz, McLoughlin:2022ljp}.

\paragraph{Gluons.} A tree level gluon amplitude in gauge theory can be decomposed as
\be\label{colored}
A(1^{a_1}\,2^{a_2}\,3^{a_3}\cdots n^{a_n}) = \sum_{\sigma\in S_{n-2}} C^{a_1a_2a_{\sigma_3}\dots a_{\sigma_n}}A[1\,2\,\sigma_3\cdots\sigma_n]\,.
\ee
Here, the $a_i$ are Lie algebra indices for the gauge group. $A[1\,2\,\sigma_3\cdots\sigma_n]$ denotes a color-ordered amplitude in the ordering associated to a permutation $\sigma$ over $n-2$ legs. It carries a factor of the momentum conserving delta function $\delta^4(p_1+p_2+\cdots+p_n)$. The coefficients $C^{a_1a_2a_{\sigma_3}\dots a_{\sigma_n}}$ are color factors that can be written in a variety of bases. To make contact with a CFT interpretation, it is convenient to work in the Del Duca-Dixon-Maltoni (DDM) basis \cite{DelDuca:1999rs}. DDM take the color factors to be
\be\label{ddm}
C^{a_1a_2a_3\dots a_n} = f^{a_2a_3}{}_{b_1}f^{b_1a_4}{}_{b_2}\cdots f^{b_{n-4}a_{n-1}}{}_{b_{n-3}} f^{b_{n-3}a_na_1}\,,
\ee
where $f^{ab}{}_c$ are the gauge group's structure constants, and Lie algebra indices are raised using the Killing form as usual.

Inverse soft recursion constructs $n$-point MHV amplitudes by attaching soft factors to lower multiplicity MHV amplitudes. For color-ordered gluon amplitudes $A[1\,2\,3\cdots n]$, this recurses over smaller color-ordered amplitudes while preserving the ordering of the labels. For concreteness, we will take gluon $1$ to be positive helicity. The two negative helicity gluons need not be explicitly specified for the recursion to work, but the reader may feel free to take them to be gluons $r$ and $s$. 

Let $p_i^{\al\dal}=\lambda_i^\al\tilde\lambda_i^{\dal}$, $\al=1,2,\dal=\dot1,\dot2$, denote the null momenta of the gluons, expressed in terms of standard spinor-helicity variables \cite{Elvang:2013cua}. The inverse soft recursion for the color-ordered MHV amplitude only involves a single term on its right hand side:
\be\label{isr}
A[1_+\,2\,3\cdots n] = \frac{\la n2\ra}{\la n1\ra\la12\ra}\,A[\hat 2\cdots n-1\,\hat n]\,,
\ee
where the subscript on $1_+$ singles it out as a positive helicity particle for clarity. The hatted particles have momenta $\hat p_2^{\al\dal} = \lambda_2^\al\hat{\tilde\lambda}_2^{\dal}$ and $\hat p_n^{\al\dal} = \lambda_n^\al\hat{\tilde\lambda}_n^{\dal}$, with the hatted spinor-helicity variables taking the deformed values
\be
\hat{\tilde\lambda}_2 = \tilde\lambda_2+\frac{\la n1\ra}{\la n2\ra}\,\tilde\lambda_1\,,\qquad \hat{\tilde\lambda}_n = \tilde\lambda_n + \frac{\la12\ra}{\la n2\ra}\,\tilde\lambda_1\,.
\ee
Here and in what follows, $\la ij\ra \vcentcolon= \eps_{\beta\al}\lambda_i^\al\lambda_j^\beta$ and $[ij] \vcentcolon= \eps_{\dot\beta\dal}\tilde\lambda_i^{\dal}\tilde\lambda_j^{\dot\beta}$ denote Lorentz invariant spinor contractions built out of the $\SL(2,\C)$ invariant Levi Civita symbols $\eps_{\al\beta},\eps_{\dal\dot\beta}$. An application of Schouten's identity shows that $\hat p_2 + p_3 + \cdots + \hat p_n = p_1+p_2+\cdots+p_n$, so that $A[\hat 2\cdots n-1\,\hat n]$ also automatically contains the correct $(n-1)$-point momentum conserving delta functions $\delta^4(\hat p_2+p_3+\cdots+\hat p_n)$.

This recursion can be derived from BCFW recursion \cite{Boucher-Veronneau:2011rwd}. In the past, it has been used for obtaining the subleading soft gluon theorem \cite{Casali:2014xpa}. To extend it to the color-dressed amplitude \eqref{colored}, relabel the index $\sigma_n\equiv i$ and break the sum over $\sigma\in S_{n-2}$ into a sum over $i$ times a sum over permutations $\pi\in S_{n-3}$ of the remaining $n-3$ gluons:
\be\label{colored1}
A(1^{a_1}\,2^{a_2}\,3^{a_3}\cdots n^{a_n}) = \sum_{i=3}^n\sum_{\pi\in S_{n-3}[i]} C^{a_1a_2a_{\pi_3}\dots a_{\pi_{n}}a_i}A[1\,2\,\pi_3\cdots\pi_{n}\,i]\,,
\ee
where $S_{n-3}[i]$ denotes permutations of the set $\{3,\dots,n\}-\{i\}$, so that the subscript on $\pi_a$ runs over $a\in\{3,\dots,n\}-\{i\}$ in the $i^\text{th}$ summand of the first sum. Inverse soft recursions for the summands are found by permuting the particle labels in \eqref{isr}:
\be
A[1_+\,2\,\pi_3\cdots\pi_{n}\,i] = \frac{\la i2\ra}{\la i1\ra\la12\ra}\,A[\hat 2\,\pi_3\cdots\pi_{n}\,\hat i]\,.
\ee
From \eqref{ddm}, we also find a recursion for the DDM color factor,
\be
C^{a_1a_2a_{\pi_3}\dots a_{\pi_{n}}a_i} = -f^{a_1a_i}{}_{b}\,C^{ba_2a_{\pi_3}\dots a_{\pi_{n}}},
\ee
where $C^{ba_2a_{\sigma_3}\dots a_{\sigma_{n-1}}}$ is an $(n-1)$-gluon color factor.

Plugging these into \eqref{colored} and using cyclic symmetry $A[\hat 2\,\pi_3\cdots \pi_{n}\,\hat i] = A[\hat i\,\hat 2\,\pi_3\cdots \pi_n]$ gives a recursion for the color-dressed MHV amplitude\footnote{See \cite{Duhr:2006iq} for a similar treatment of tree level BCFW recursion in full generality.}
\be\label{gluonrec}
A(1_+^{a_1}\,2^{a_2}\cdots n^{a_n}) = -\sum_{i=3}^n \frac{\la i2\ra}{\la i1\ra\la12\ra}\,T_i^{a_1}A(\hat 2^{a_2}\cdots\hat i^{a_i}\cdots n^{a_n})\,,
\ee
where, for clarity, we repeat the definitions of the hatted particles' spinor-helicity data:
\be\label{hatted}
\begin{aligned}
    \hat\lambda_2 &= \lambda_2\,,&\quad\hat{\tilde\lambda}_2 &= \tilde\lambda_2+\frac{\la i1\ra}{\la i2\ra}\,\tilde\lambda_1,\\
    \hat\lambda_i &= \lambda_i\,,&\quad\hat{\tilde\lambda}_i &= \tilde\lambda_i + \frac{\la12\ra}{\la i2\ra}\,\tilde\lambda_1\,.
\end{aligned}
\ee
The momenta of the other particles remain undeformed. We have also introduced an operator $T^a_i$ that rotates the Lie algebra indices as
\be\label{Ta}
T_i^{a_1}A(\hat 2^{a_2}\cdots\hat i^{a_i}\cdots n^{a_n}) \vcentcolon= f^{a_1a_i}{}_b\,A(\hat 2^{a_2}\cdots\hat i^{b}\cdots n^{a_n})\,.
\ee
This is just a transformation in the adjoint representation.

\paragraph{Gravitons.} The inverse soft recursion for graviton amplitudes can also be derived from BCFW recursion. It was used in \cite{Cachazo:2014fwa} to derive the subleading and sub-subleading soft graviton theorems.

Let $M(1\,2\cdots n)$ denote an $n$-graviton amplitude. Take it to be an MHV amplitude with graviton $1$ being positive helicity. Then it satisfies the inverse soft recursion
\be\label{gravrec}
M(1_+\, 2 \cdots  n) = \sum_{i =3}^n \frac{[1i]}{\la1i\ra} \frac{\la i2\ra^2}{\la12\ra^2}\,M(\hat{2}\cdots \hat{i} \cdots n) \, .
\ee
The hatted gravitons have the same spinor-helicity variables \eqref{hatted} as the hatted gluons. There is no color-ordering to be dealt with in this case. The color factors $f^{a_1a_i}{}_b$ are instead replaced by kinematic factors $[1i]/\la1i\ra$, reflecting color-kinematics duality. This fact is also known to translate to a color-kinematics duality at the level of celestial OPEs \cite{Monteiro:2022lwm}.

\bigskip

If one goes beyond the MHV sector, such recursion relations involve further terms on their right hand sides \cite{Elvang:2013cua}. These terms generally contain multiparticle factorization poles. They cannot be interpreted as poles in the correlators of a local vertex algebra and also break the usual notions of associativity of the celestial OPE \cite{Ball:2022bgg}. We leave the study of these more involved terms to the future, which might very well require a generalization like non-local vertex algebras \cite{Bakalov2002FieldA}.

\section{MHV Gluon OPE}
\label{sec:gluon}

\subsection{Collinear expansion}

Spinor-helicity variables $\lambda_i^\al,\tilde\lambda_i^{\dal}$ are only defined up to the scalings $(\lambda_i,\tilde\lambda_i)\sim(t_i\lambda_i,t_i^{-1}\tilde\lambda_i)$ for $t_i\in\C^*$. This is known as little group scaling. Using this freedom, we choose to fix
\be\label{lgfix}
\lambda_i^\al=\begin{pmatrix}1\\z_i\end{pmatrix}\,,\qquad\tilde\lambda_i^{\dal} = \omega_i\begin{pmatrix}1\\\bz_i\end{pmatrix}\,,\qquad z_i,\bz_i\in\C\,,\;\omega_i\in\C^*\,.
\ee
$z_i,\bz_i$ are generally taken to be complex and independent of each other. They are complex conjugates only in Lorentzian signature where they act as coordinates on the celestial sphere. With this fixing of the scalings, one finds
\be
\la ij\ra = z_{ij}\,,\qquad[ij]=\omega_i\omega_j\bz_{ij}\,,
\ee
where $z_{ij}\equiv z_i-z_j$ and $\bz_{ij}\equiv\bz_i-\bz_j$. This choice of little group fixing is the most efficient when working with a holomorphic collinear expansion, i.e., an expansion in small $z_{ij}$ while keeping $[ij]$ arbitrary. While studying the twistorial origin of celestial OPEs, it was applied by Costello and Paquette to great success in \cite{Costello:2022wso,Costello:2022upu}.

Let us now return to the recursion relations. The main observation underpinning our derivation of the celestial OPE is that the hatted spinor-helicity variables \eqref{hatted} can be written in the suggestive form
\be
\begin{split}
    \hat{\tilde\lambda}_2 &= \tilde\lambda_2 + \frac{z_{i1}}{z_{i2}}\tilde\lambda_1 = \tilde\lambda_1+\tilde\lambda_2 - \frac{z_{12}}{z_{i2}}\,\tilde\lambda_1\,,\\
    \hat{\tilde\lambda}_i &= \tilde\lambda_i + \frac{z_{12}}{z_{i2}}\,\tilde\lambda_1\,.
\end{split}
\ee
As a result, the inverse soft recursion \eqref{gluonrec} for the gluon MHV amplitude can be recast in terms of exponentiations of linear shifts of $\tilde\lambda_2$ and $\tilde\lambda_i$:
\be\label{gluonrec1}
A(1_+^{a_1}\,2^{a_2}\cdots n^{a_n}) = -\sum_{i=3}^n \frac{1}{z_{12}}\biggl(1-\frac{z_{12}}{z_{i2}}\biggr)^{-1} \exp\biggl\{\frac{z_{12}}{z_{i2}}\bigl([1\p_i]-[1\p_2]\bigr)\biggr\}\, T^{a_1}_iA(\boldsymbol{2}^{a_2}\cdots n^{a_n})  
\ee
where we have introduced the abbreviation 
\be
[i\p_j]\equiv\tilde\lambda_i^{\dal}\,\frac{\p}{\p\tilde\lambda_j^{\dal}} = \frac{\omega_i}{\omega_j}\left(\bz_{ij}\frac{\p}{\p\bz_j}+\omega_j\frac{\p}{\p\omega_j}\right)\,.
\ee
On the right hand side of \eqref{gluonrec1}, particles $3,\dots,n$ now carry the original undeformed momenta $p_3,\dots,p_n$. On the other hand, particle $2$ is depicted in bold to signify that it carries the ``effective'' momentum
\be
\mathbf{p}_2^{\al\dal} = \lambda_2^\al(\tilde\lambda_1+\tilde\lambda_2)^{\dal}\,.
\ee
Because $\lambda_j^1=1\;\forall\;j=1,\dots,n$ in our little group fixing, if needed, $\mathbf{p}_2^{\al\dal}$ can also be written in a little group invariant form by replacing $\tilde\lambda_1^{\dal}\mapsto(\lambda_1^1/\lambda_2^1)\tilde\lambda_1^{\dal}$. 

These steps have brought the recursion to a form that is ripe for holomorphic collinear expansion. Taylor expanding \eqref{gluonrec1} around $z_{12}=0$ yields 
\be
A(1_+^{a_1}\,2^{a_2}\cdots n^{a_n}) = -\sum_{i=3}^n\sum_{p=0}^\infty\frac{z_{12}^{p-1}}{z_{i2}^p}\sum_{q=0}^p\frac{([1\p_i]-[1\p_2])^q}{q!}\,T^{a_1}_iA(\boldsymbol{2}^{a_2}\cdots n^{a_n})\,.
\ee
Further expanding the factors of $([1\p_i]-[1\p_2])^q$ using the binomial theorem, one finds the all-order collinear expansion of the tree level MHV gluon amplitude,
\be\label{gluoncoll}
\boxed{A(1_+^{a_1}\,2^{a_2}\cdots n^{a_n}) = -\sum_{p=0}^\infty z_{12}^{p-1}\sum_{q=0}^p\sum_{r=0}^q\frac{(-[1\p_2])^{q-r}}{r!(q-r)!}\sum_{i=3}^n\frac{[1\p_i]^r}{z_{i2}^p}\,T^{a_1}_iA(\boldsymbol{2}^{a_2}\cdots n^{a_n})\,.}
\ee
This holds regardless of the helicity of gluon 2. 


\subsection{Celestial OPE of momentum eigenstates}

Celestial holography posits the existence of two dimensional operators $O^{a_i}_{s_i}(z_i,\tilde\lambda_i)$ in the celestial CFT that are dual to gluon momentum eigenstates. The subscripts denote helicities $s_i\in\{\pm1\}$. Correlators of such operators are conjectured to reproduce scattering amplitudes in the bulk,
\be\label{duality}
A(1^{a_1}\,2^{a_2}\cdots n^{a_n}) = \left\la O_{s_1}^{a_1}(z_1,\tilde\lambda_1)\,O_{s_2}^{a_2}(z_2,\tilde\lambda_2)\cdots O_{s_n}^{a_n}(z_n,\tilde\lambda_n)\right\ra\,.
\ee
In particular, one reads off conjectural operator product expansions of the dual operators by demanding that the OPE expansions of the correlators reproduce collinear limits of scattering amplitudes. The resulting operator products are known as \emph{celestial OPEs}.

\paragraph{Soft gluon currents.} In the MHV sector, we can use this logic to read off gluon celestial OPEs to all orders from \eqref{gluoncoll}. To do so, we need to introduce positive helicity soft gluon currents. The soft expansion of a positive helicity hard gluon $O_+^a(z,\tilde\lambda)$ with $\tilde\lambda^{\dal}=\omega(1,\bz)$ is a  Taylor expansion in the energy $\omega$.\footnote{Our soft expansion starts at $\omega^0$ in this little group fixing instead of with the usual Weinberg soft pole $\omega^{-1}$ that one encounters for the choice $\lambda^\al=\sqrt{\omega}(1,z)$, $\tilde\lambda^{\dal}=\sqrt{\omega}(1,\bz)$. This is just a convenient convention.} Equivalently, it can be represented as a double Taylor expansion in $\tilde\lambda^{\dot1},\tilde\lambda^{\dot2}$,
\be\label{eq:soft_def}
O_+^a(z,\tilde\lambda) = \sum_{k,\ell=0}^\infty\frac{(\tilde\lambda^{\dot1})^k(\tilde\lambda^{\dot2})^\ell}{k!\,\ell!}\,J^a[k,\ell](z)\,.
\ee
The Taylor coefficients $J^a[k,\ell](z)$ are chiral currents known as \emph{soft gluon currents}. Under M\"obius transformations of $z$, they transform as chiral conformal primaries with conformal weights $h=1-\frac{k+\ell}{2}$. Similarly, under $\SL(2,\C)$ rotations of $\tilde\lambda^{\dal}$, the currents at fixed $k+\ell$ span a spin $\frac{k+\ell}{2}$ representation of $\SL(2,\C)$.

Celestial OPEs of these soft currents were determined in \cite{Guevara:2021abz,Strominger:2021lvk}. Strominger further recognized in \cite{Strominger:2021lvk} that their modes generate the loop algebra of holomorphic maps $\C^2\to\mathfrak{g}$, where $\mathfrak{g}$ is the gauge group's Lie algebra. Hence, this loop algebra was established to be a symmetry algebra of the celestial CFT dual to gauge theory. For the purposes of this work, we are adopting the simplifying notation of \cite{Costello:2022wso} for these currents.\footnote{Our currents are related to Strominger's $S$-algebra currents as $S^{p,a}_m(z) = J^a[p-1+m,p-1-m](z)$, where $p\in\{1,\frac{3}{2},2,\frac{5}{2},\dots\}$ and $|m|\leq p-1$.}

The leading singularity in \eqref{gluoncoll} allows us to extract the singular part of the OPE. To see this, begin by truncating \eqref{gluoncoll} to the singular order,
\be
A(1_+^{a_1}\,2^{a_2}\cdots n^{a_n}) = -\frac{1}{z_{12}}\sum_{i=3}^nT^{a_1}_iA(\boldsymbol{2}^{a_2}\cdots n^{a_n}) + \rO(z_{12}^0)\,.
\ee
Global color conservation for the $(n-1)$-point amplitude $A(\boldsymbol{2}^{a_2}\cdots n^{a_n})$ gives the identity
\be
\sum_{i=3}^nT^{a_1}_iA(\boldsymbol{2}^{a_2}\cdots n^{a_n}) = -T^{a_1}_2A(\boldsymbol{2}^{a_2}\cdots n^{a_n}) = -f^{a_1a_2}{}_b\,A(\boldsymbol{2}^{b}\cdots n^{a_n})\,.
\ee
This is the conservation law associated to the leading soft gluon theorem \cite{He:2015zea}. Using this, we obtain the singular part of the collinear expansion,
\be
A(1_+^{a_1}\,2^{a_2}\cdots n^{a_n}) = \frac{f^{a_1a_2}{}_b}{z_{12}}\,A(\boldsymbol{2}^{b}\cdots n^{a_n}) + \rO(z_{12}^0)\,.
\ee
This provides the standard conjecture for the holomorphic hard gluon celestial OPE at order $1/z_{12}$ and all orders in $\bz_{12}$ \cite{Fan:2019emx,Pate:2019lpp, Guevara:2021abz,Himwich:2021dau}
\be\label{hardgluonope}
O^{a_1}_+(z_1,\tilde\lambda_1)\,O^{a_2}_{s_2}(z_2,\tilde\lambda_2) \sim \frac{f^{a_1a_2}{}_b}{z_{12}}\;O^b_{s_2}(z_2,\tilde\lambda_1+\tilde\lambda_2)\,,\qquad s_2\in\{\pm1\}\,.
\ee
Thankfully, singularities in $\bz_{12}$ do not come into play as long as we stay within the MHV sector, giving us a perfect toy model to work with.

Choosing $s_2=+1$ and expanding in $\tilde\lambda_1^{\dal},\tilde\lambda_2^{\dal}$, one can read off the current algebra of soft gluons,
\be
J^a[k,\ell](z)\,J^b[m,n](w) \sim \frac{f^{ab}{}_c}{z-w}\;J^c[k+m,\ell+n](w)\,.
\ee
This has vanishing level (at least at tree level in the bulk). As mentioned before, this is the loop algebra of $\text{Maps}(\C^2,\mathfrak{g})$ found in \cite{Strominger:2021lvk}.\footnote{Holomorphic maps $\C^2\to\mathfrak{g}$ are generated by the monomials $v_1^kv_2^\ell\mathfrak{t}^a$, where $v_1,v_2$ are complex coordinates on $\C^2$ and $\mathfrak{t}^a$ are generators of $\mathfrak{g}$ with Lie bracket $[\mathfrak{t}^a,\mathfrak{t}^b]=f^{ab}{}_c\mathfrak{t}^c$. The set of such holomorphic maps has the natural Lie algebra structure $[v_1^kv_2^\ell\mathfrak{t}^a,v_1^mv_2^n\mathfrak{t}^b] = v_1^{k+m}v_2^{\ell+n}f^{ab}{}_c\mathfrak{t}^c$. The soft gluon OPE is seen to be the loop algebra based on this Lie algebra.} It originates in the symmetries of the Penrose-Ward transform as elaborated in \cite{Costello:2022wso}.

It is also useful to define certain currents that fall between the hard and soft gluons:
\be\label{eq:J_lamb_def}
J^a[r](z,\tilde\lambda) = \sum_{k=0}^r\frac{(\tilde\lambda^{\dot1})^k(\tilde\lambda^{\dot2})^{r-k}}{k!(r-k)!}\, J^a[k,r-k](z)\,,\qquad r\in\Z_{\geq0}\,.
\ee
These act as generating functions of the soft gluon currents. In terms of these, the positive helicity hard gluons admit the clean decomposition
\be
O_+^a(z,\tilde\lambda) = \sum_{r=0}^\infty J^a[r](z,\tilde\lambda)\,.
\ee
Because $J^a[r](z,\tilde\lambda)$ is homogeneous of weight $r$ in $\tilde\lambda^{\dal}=\omega(1,\bz)$, the $r^\text{th}$ term of this expansion is just $\omega^r$ times the $r^\text{th}$ subleading soft gluon. 

The OPE between soft and hard gluons can be compactly expressed as
\be\label{softhard}
J^{a_1}[r](z_1,\tilde\lambda_1)\,O_{s_i}^{a_i}(z_i,\tilde\lambda_i) \sim \frac{f^{a_1a_i}{}_b}{z_{1i}}\,\frac{[1\p_i]^r}{r!}\,O_{s_i}^{b}(z_i,\tilde\lambda_i)\,.
\ee
This is a momentum space version of the boost eigenstate soft-hard OPEs found in \cite{Himwich:2021dau}. It is obtained by relabeling $2\mapsto i$ in \eqref{hardgluonope}, expanding $O^b_{s_i}(z_i,\tilde\lambda_1+\tilde\lambda_i) = \e^{[1\p_i]}O^b_{s_i}(z_i,\tilde\lambda_i)$ in powers of $[1\p_i]$, then comparing terms of homogeneity $r$ in $\tilde\lambda_1^{\dal}$ on both sides. The operator $[1\p_i]\equiv\tilde\lambda_1\cdot\p_{\tilde\lambda_i}$ differentiates the $\tilde\lambda_i^{\dal}$ dependence of the hard gluon. The simple expression \eqref{softhard} is the main upshot of using the generating functions $J^{a}[r]$. OPEs between soft gluon currents $J^{a_1}[k,\ell](z_1)$ and hard gluon operators can be extracted by equating the coefficients of $(\tilde\lambda^{\dot1}_1)^k(\tilde\lambda^{\dot2}_1)^\ell$.

\paragraph{Soft gluon descendants.} In the MHV sector, one does not encounter singularities in $\bz_{12}$. Complexifying $z_i,\bz_i$ (or working in split signature), one can therefore continue to treat the soft gluons $J^a[r](z,\tilde\lambda)$ as chiral currents living on the sphere coordinatized by $z$. That is, we are treating $\tilde\lambda^{\dal}$ as an operator label rather than a coordinate on the Riemann sphere on which the CCFT is defined. In particular, we can define $J^a[r]$ descendants as the $\tilde\lambda$-dependent operators
\be\label{eq:Jp_def}
J^{a_1}_{-p}[r](\tilde\lambda_1)O^{a_2}_{s_2}(z_2,\tilde\lambda_2) \vcentcolon= \oint_{|z_{12}|=\veps}\frac{\d z_1}{2\pi\im}\,\frac{1}{z_{12}^{p}}\,J^{a_1}[r](z_1,\tilde\lambda_1)\,O^{a_2}_{s_2}(z_2,\tilde\lambda_2),
\ee
where $\veps\ll1$ and $p\geq0$. The integrand of the contour integral is to be understood as the radially ordered operator product as is standard in 2d CFT.

Suppose one inserts such a descendant in a correlation function of hard gluon operators,
\begin{multline}\label{descorr0}
    \left\la J^{a_1}_{-p}[r](\tilde\lambda_1)O^{a_2}_{s_2}(z_2,\tilde\lambda_2)\prod_{i=3}^n O^{a_i}_{s_i}(z_i,\tilde\lambda_i)\right\ra \\
    = \oint_{|z_{12}|=\veps}\frac{\d z_1}{2\pi\im}\,\frac{1}{z_{12}^{p}}\left\la J^{a_1}[r](z_1,\tilde\lambda_1)\,O^{a_2}_{s_2}(z_2,\tilde\lambda_2)\prod_{i=3}^n O^{a_i}_{s_i}(z_i,\tilde\lambda_i)\right\ra\,.
\end{multline}
Since the current $J^a[r](z,\tilde\lambda)$ transforms as a primary of weight $h=1-r/2$ under M\"obius transformations of $z$, it has a pole of order $r-2$ at $z=\infty$. Because of this, the correlator on the right can also a priori have a pole of order $r-2$ at $z_1=\infty$.\footnote{We thank Elizabeth Himwich as well as one of the referees of this paper for emphasizing this subtlety.} 

In a garden variety CFT, we would have expected such a pole to drop out entirely and only two-particle singularities at $z_{ij}=0$ to survive. We prove in appendix \ref{app:sing} the weaker result that the soft-hard amplitude that is supposed to equal the correlator on the right of \eqref{descorr0} contains a pole at infinity that is indeed of order at most $r-2$. This is all we need for what follows. It might still be the case that the residue at this pole vanishes, so that the pole may be spurious, but we leave this analysis to the future. In fact, such poles are expected to be related to the absence of universal soft theorems beyond the subleading order.

All in all, because correlators of $J^{a_1}[r](z_1,\tilde\lambda_1)$ only have a pole of order $r-2$ or less at $z_1=\infty$, the integrand of the contour integral in \eqref{descorr0} definitely has no singularities at $z_1=\infty$ for $0\leq r\leq p$.\footnote{Keep in mind that $\d z_1$ itself has a pole of order $2$ while $z_{12}^{-p}$ has a zero of order $p$ at infinity.} In this range of $r$, the contour of integration can be deformed to surround all the other poles at $z_1=z_i$, $i=3,\dots,n$, instead of the original pole at $z_1=z_2$. This converts the right hand side to
\be\label{contourdef}
-\sum_{i=3}^n\oint_{|z_{1i}|=\veps}\frac{\d z_1}{2\pi\im}\,\frac{1}{z_{12}^{p}}\left\la J^{a_1}[r](z_1,\tilde\lambda_1)\,O^{a_2}_{s_2}(z_2,\tilde\lambda_2)\prod_{j=3}^n O^{a_j}_{s_j}(z_j,\tilde\lambda_j)\right\ra\,.
\ee
The pole at $z_{1i}=0$ is a simple pole with residue fixed by the soft-hard OPE \eqref{softhard}. Hence, for $r\leq p$, the descendant correlator can be expressed purely in terms of hard gluon correlators:
\begin{align}
    \left\la J^{a_1}_{-p}[r](\tilde\lambda_1)O^{a_2}_{s_2}(z_2,\tilde\lambda_2)\prod_{i=3}^n O^{a_i}_{s_i}(z_i,\tilde\lambda_i)\right\ra &= -\frac{1}{r!}\sum_{i=3}^n\frac{f^{a_1a_i}{}_b}{z_{i2}^p}\left\la[1\p_i]^rO^{b}_{s_i}(z_i,\tilde\lambda_i)\prod_{\substack{j=2\\j\neq i}}^n O^{a_j}_{s_j}(z_j,\tilde\lambda_j)\right\ra\nonumber\\
    &= -\frac{1}{r!}\sum_{i=3}^n\frac{[1\p_i]^r}{z_{i2}^p}\,T^{a_1}_iA(2^{a_2}\,3^{a_3}\cdots n^{a_n})\,.\label{descorr}
\end{align}
To get the second line, we have used the duality statement \eqref{duality} as well as the definition \eqref{Ta} of the adjoint action $T^{a_1}_i$. Luckily, we will only require descendant correlators for the range $r\leq p$ when extracting the OPE.

\paragraph{All-order gluon OPE.} Having defined the soft gluon descendants and their correlators, it becomes straightforward to extract the all-order celestial OPE from the all-order collinear expansion \eqref{gluoncoll}. 

Using the duality statement \eqref{duality} and the descendant correlator \eqref{descorr} -- taken after a shift $\tilde\lambda_2\mapsto\tilde\lambda_1+\tilde\lambda_2$ -- we get the all-order OPE expansion for CCFT correlators in the MHV sector,
\begin{multline}
    \left\la O^{a_1}_+(z_1,\tilde\lambda_1)\,O^{a_2}_{s_2}(z_2,\tilde\lambda_2)\prod_{i=3}^n O^{a_i}_{s_i}(z_i,\tilde\lambda_i)\right\ra\\
    = \sum_{p=0}^\infty z_{12}^{p-1}\sum_{q=0}^p\sum_{r=0}^q\frac{(-[1\p_2])^{q-r}}{(q-r)!}\left\la J^{a_1}_{-p}[r](\tilde\lambda_1)O^{a_2}_{s_2}(z_2,\tilde\lambda_1+\tilde\lambda_2)\prod_{i=3}^n O^{a_i}_{s_i}(z_i,\tilde\lambda_i)\right\ra\,.
\end{multline}
As foreshadowed, the index $r$ only runs up to $p$. So we are happily in the range in which the contour deformation used to obtain \eqref{contourdef} holds. 

This relation is equivalent to giving the following OPE to gluons $1$ and $2$:
\be
O^{a_1}_+(z_1,\tilde\lambda_1)\,O^{a_2}_{s_2}(z_2,\tilde\lambda_2) = \sum_{p=0}^\infty z_{12}^{p-1}\sum_{q=0}^p\sum_{r=0}^q\frac{(-[1\p_2])^{q-r}}{(q-r)!}\,J^{a_1}_{-p}[r](\tilde\lambda_1)O^{a_2}_{s_2}(z_2,\tilde\lambda_1+\tilde\lambda_2)\,.
\ee
It can be further cleaned up by exchanging the sums over $q$ and $r$,
\be\label{eq:gluonopebox}
\boxed{O^{a_1}_+(z_1,\tilde\lambda_1)\,O^{a_2}_{s_2}(z_2,\tilde\lambda_2) = \sum_{p=0}^\infty z_{12}^{p-1}\sum_{r=0}^p\sum_{q=0}^{p-r}\frac{(-[1\p_2])^{q}}{q!}\,J^{a_1}_{-p}[r](\tilde\lambda_1)O^{a_2}_{s_2}(z_2,\tilde\lambda_1+\tilde\lambda_2)\,.}
\ee
This is our result for the momentum space all-order celestial OPE of two gluons participating in an MHV amplitude. It holds for either value $s_2=\pm1$ of the helicity of the second gluon. To get a feel for this, it is useful to write out the first few terms of the sum explicitly,
\begin{align}
    &O^{a_1}_+(z_1,\tilde\lambda_1)\,O^{a_2}_{s_2}(z_2,\tilde\lambda_2) \nonumber\\
    =& \frac{1}{z_{12}}\,J_0^{a_1}[0](\tilde\lambda_1)O^{a_2}_{s_2}(z_2,\tilde\lambda_1+\tilde\lambda_2) +\Bigl\{\bigl(1-[1\p_2]\bigr)\,J^{a_1}_{-1}[0](\tilde\lambda_1) + J^{a_1}_{-1}[1](\tilde\lambda_1)\Bigr\}\,O^{a_2}_{s_2}(z_2,\tilde\lambda_1+\tilde\lambda_2)\nonumber\\
    &+z_{12}\,\biggl\{\bigg(1-[1\p_2]+\frac{[1\p_2]^2}{2!}\bigg)\,J^{a_1}_{-2}[0](\tilde\lambda_1) + \bigl(1-[1\p_2]\bigr)\,J^{a_1}_{-2}[1](\tilde\lambda_1) + J^{a_1}_{-2}[2](\tilde\lambda_1)\biggr\}\,O^{a_2}_{s_2}(z_2,\tilde\lambda_1+\tilde\lambda_2)\nonumber\\
    &+\rO(z_{12}^2)\,.
\end{align}
The leading term is of course just the leading ``descendant'' $J_0^{a_1}[0](\tilde\lambda_1)O^{a_2}_{s_2}(z_2,\tilde\lambda_1+\tilde\lambda_2) = f^{a_1a_2}{}_b\,O^{a_2}_{s_2}(z_2,\tilde\lambda_1+\tilde\lambda_2)$. The first subleading term receives contributions from descendants of both the leading and subleading soft gluon symmetries. Further terms receive contributions from soft gluons that sit at higher orders in the soft expansion.

As expected, \eqref{eq:gluonopebox} matches the all-order OPE obtained from the explicit vertex algebra realization coming from twistor string theory \cite{Adamo:2022wjo}. But even though finding such realizations is the ultimate goal of the celestial holography program, computations using explicit vertex operators can be cumbersome and depend intimately on the specific vertex algebra being studied. On the other hand, our bottom-up derivation of the OPE from on-shell recursion is more transparent and relatively more universal. 

It also allows us to conclude that \emph{inverse soft recursion for MHV amplitudes is dual to OPE recursion for CCFT correlators}. In this sense, on-shell recursion relations can potentially emerge from flat space holography. Since inverse soft recursion is just BCFW recursion in disguise, we hope that a similar statement might hold for BCFW recursion beyond the MHV sector, but this is left to future work.


\subsection{Celestial OPE of boost eigenstates}\label{sec:gluonboost}

The celestial OPE of boost eigenstates can be found by Mellin transforming the momentum space OPE. This calculation was already performed in \cite{Adamo:2022wjo}, so we only repeat the main ideas.

Suppose we restrict to split signature kinematics following the lines of \cite{Atanasov:2021cje}. Decompose the dotted spinor helicity variable associated to a momentum eigenstate operator $O^a_s(z,\tilde\lambda)$ as
\be
\tilde\lambda^{\dal} = \veps\,\omega\begin{pmatrix}
    1\\\bz
\end{pmatrix}\,,\quad \veps\in\{\pm1\}\,,\;\omega\in(0,\infty)\,,
\ee
with $z,\bz$ now being real and independent. Let $\Delta\in\C$ and $h=(\Delta+s)/2$, $\bh = (\Delta-s)/2$. Celestial operators dual to boost eigenstates of weight $\Delta$ and spin $s$ are found by Mellin transforming the momentum eigenstate operators,
\be\label{eq:mellin}
O_{\Delta,s}^{\veps,a}(z,\bz) = \int_0^\infty\frac{\d\omega}{\omega}\,\omega^{2\bh}\,O^a_{s}(z,\tilde\lambda)\,.
\ee
Under $\SL(2,\R)\times\SL(2,\R)$ transformations of $z,\bz$, these operators transform as conformal primaries of weights $(h,\bh)$. So the basis of boost eigenstates is also known as a conformal basis, first constructed in \cite{Pasterski:2017kqt}.

Define the conformal primary positive helicity soft gluons
\be\label{eq:softgluon}
J^a[r](z,\bz) \vcentcolon= \frac{J^a[r](z,\tilde\lambda)}{(\veps\omega)^r} = \sum_{k=0}^r\frac{\bz^{r-k}J^a[k,r-k](z)}{k!(r-k)!}\,.
\ee
Equivalently, one can define $J^a[r]$ through residues of the conformal primary gluon operators \cite{Guevara:2021abz}. The descendants appearing in their OPE with boost eigenstates are also defined by the usual contour integrals
\be
J^a_{-p}[r](\bz)O^{\veps,b}_{\Delta,s}(w,\bar w) = \oint\frac{\d z}{2\pi\im}\,\frac{1}{(z-w)^p}\,J^a[r](z,\bz)\,O^{\veps,b}_{\Delta,s}(w,\bar w) \,,
\ee
where the contour surrounds the pole at $z=w$. Let's see how to express the celestial OPE of boost eigenstates in terms of these descendants.

Mellin transforming both sides of \eqref{eq:gluonopebox} leads to the following set of integrals:
\begin{equation}\label{eq:mellingluon}
    \begin{aligned}
        & O^{\veps_1,a_1}_{\Delta_1,+}(z_1,\bz_1)\,O^{\veps_2,a_2}_{\Delta_2,s_2}(z_2,\bz_2) \\
        =& \sum_{p=0}^\infty z_{12}^{p-1}\sum_{r=0}^p\sum_{q=0}^{p-r}\sum_{m=0}^\infty\frac{(-1)^q}{q!m!}\int_0^\infty\d\omega_1\,\omega_1^{2\bh_1-1}\int_0^\infty\d\omega_2\,\omega_2^{2\bh_2-1} (\veps_1\omega_1)^{r+m}\\
        & \times \bigg(\frac{\veps_1\omega_1}{\veps_2\omega_2}\big(\bz_{12}\dbar_2+\omega_2\p_{\omega_2}\big)\bigg)^q\bz_{12}^m\dbar_2^mJ^{a_1}_{-p}[r](\bz_1)O^{a_2}_{s_2}\bigg(z_2,\frac{(\veps_1\omega_1+\veps_2\omega_2)}{\veps_2\omega_2}\tilde\lambda_2\bigg),
    \end{aligned}
\end{equation}
where $(h_i,\bh_i) = (\frac{\Delta_i+s_i}{2},\frac{\Delta_i-s_i}{2})$, and $\dbar_i\equiv\p/\p\bz_i$. The sum over $m$ is just the Taylor expansion in $\bz_{12}$ of the momentum eigenstate on the right of \eqref{eq:gluonopebox}. 


\paragraph{$\veps_1=\veps_2$.} In this case, we can introduce the change of integration variables
\be
\omega_1 = t\omega\,,\quad\omega_2=(1-t)\omega\,,\quad t\in(0,1)\,,\quad\omega\in(0,\infty)\,,
\ee
under which the integral \eqref{eq:mellingluon} turns into (after some math):
    \begin{align}\label{eq:gluonOPEmid2}
        & O^{\veps_1,a_1}_{\Delta_1,+}(z_1,\bz_1)\,O^{\veps_2,a_2}_{\Delta_2,s_2}(z_2,\bz_2) \\
        =& \sum_{p,m=0}^\infty z_{12}^{p-1} \sum_{r=0}^p \sum_{q=0}^{p-r} \sum_{q'=0}^{q} \sum_{s=0}^{q'} \int_0^\infty\d\omega\,\omega^{2\bh_1+2\bh_2-1+r} \int_{0}^{1} \d t\, t^{2\bh_1-1+r+q+m} (1-t)^{2\bh_2-q-1}\nonumber\\
        & \times \frac{\varepsilon_1^{r} (-1)^{q-s} \, \Gamma(-2\bh_2+q-q'+1) \,\bz_{12}^{q'+m-s}}{(q-q')! \,s!\, (q'-s)! (m-s)! \, \Gamma(-2\hb_2+1)} \dbar_2^{q'+m-s} J_{-p}^{a_1}[r](\bz_1) O^{a_2}_{s_2}\bigg( z_2, \frac{(\veps_1\omega_1+\veps_2\omega_2)}{\veps_2\omega_2}\tilde\lambda_2 \bigg) \,. \nonumber
    \end{align}
Performing the Mellin transform as well as the summation we finally get:
\begin{equation}\label{eq:gluonOPE}
    \boxed{\begin{aligned}
        O^{\veps_1,a_1}_{\Delta_1,+}(z_1,\bz_1)\,O^{\veps_2,a_2}_{\Delta_2,s_2}(z_2,\bz_2) =& \sum_{p,\bar{m}=0}^\infty \sum_{r=0}^p \frac{\varepsilon_1^{r}}{\bar{m}!} B(2\hb_1+r+\bar{m},2\hb_2) \frac{\Gamma(2\hb_1+p+1)}{(p-r)!\, \Gamma(2\hb_1+r+1)} \\
        & \times z_{12}^{p-1} \, \bz_{12}^{\bar{m}} \dbar_2^{\bar{m}} \, J_{-p}^{a_1}[r](\bz_1) O^{\varepsilon_1,a_2}_{\Delta_1+\Delta_2+r-1,s_2}(z_2,\bz_2).
    \end{aligned}}
\end{equation}
We have put the derivation details in Appendix~\ref{app:computeOPE}. 

\paragraph{$\veps_1=-\veps_2$.} The OPE when $\varepsilon_1 = -\varepsilon_2$ gives a similar formula only that there are two terms produced corresponding to different ingoing/outgoing directions:
\begin{equation}\label{eq:gluonOPEdifferentep}
    \boxed{\begin{aligned}
        & O^{\veps_1,a_1}_{\Delta_1,+}(z_1,\bz_1)\,O^{\veps_2,a_2}_{\Delta_2,s_2}(z_2,\bz_2) \\
        =& \sum_{p,\bar{m}=0}^\infty \sum_{r=0}^p \frac{\varepsilon_2^{r}}{\bar{m}!} (-1)^{r+\bar{m}} B(2\hb_1+r+\bar{m},-2\hb_1 - 2\hb_2 -r-\bar{m}+1) \, \frac{\Gamma(2\hb_1+p+1)}{(p-r)!\, \Gamma(2\hb_1+r+1)} \\
        & \qquad\qquad\qquad\times z_{12}^{p-1} \, \bz_{12}^{\bar{m}} \dbar_2^{\bar{m}} \, J_{-p}^{a_1}[r](\bz_1) O^{\varepsilon_2,a_2}_{\Delta_1+\Delta_2+r-1,s_2}(z_2,\bz_2) \\
        & + \sum_{p,\bar{m}=0}^\infty \sum_{r=0}^p \frac{\varepsilon_1^{r}}{\bar{m}!} B(-2\hb_1 - 2\hb_2 -r-\bar{m}+1, 2\hb_2) \, \frac{\Gamma(2\hb_1+p+1)}{(p-r)!\, \Gamma(2\hb_1+r+1)} \\
        & \qquad\qquad\qquad\times z_{12}^{p-1} \, \bz_{12}^{\bar{m}} \dbar_2^{\bar{m}} \, J_{-p}^{a_1}[r](\bz_1) O^{\varepsilon_2,a_2}_{\Delta_1+\Delta_2+r-1,s_2}(z_1,\bz_2) \,. \\
    \end{aligned}}
\end{equation}

\section{MHV Graviton OPE}
\label{sec:graviton}


\subsection{Collinear expansion}

A similar story holds for MHV graviton scattering. In the choice of little group fixing $\lambda_i^\al=(1,z_i)$, the inverse soft recursion \eqref{gravrec} for graviton MHV amplitudes takes the form
\be
M(1_+\, 2 \cdots  n) = \sum_{i =3}^n \frac{[1i]}{z_{1i}} \frac{z_{i2}^2}{z_{12}^2}\,M(\hat{2}\cdots \hat{i} \cdots n) \,,
\ee
wherein the hatted spinor-helicity variables read
\be
\hat{\tilde\lambda}_2 = \tilde\lambda_1+\tilde\lambda_2 - \frac{z_{12}}{z_{i2}}\,\tilde\lambda_1\,,\qquad\hat{\tilde\lambda}_i = \tilde\lambda_i + \frac{z_{12}}{z_{i2}}\,\tilde\lambda_1\,.
\ee
These are the same as for the gluon case.

Once again, we can expand the $(n-1)$-graviton amplitude on the right in small $z_{12}$ by writing it in terms of exponentiations of linear shifts:
\be
M(1_+\, 2 \cdots  n) = -\sum_{i =3}^n \frac{z_{i2}[1i]}{z_{12}^2}\biggl(1-\frac{z_{12}}{z_{i2}}\biggr)^{-1} \exp\biggl\{\frac{z_{12}}{z_{i2}}\bigl([1\p_i]-[1\p_2]\bigr)\biggr\}\, M(\boldsymbol{2}\cdots n)\,.  
\ee
On the right, gravitons $3,\dots,n$ carry the undeformed null momenta $p_3,\dots,p_n$, whereas the graviton $2$ in bold carries the deformed null momentum $\mathbf{p}_2^{\al\dal}=\lambda_2^\al(\tilde\lambda_1+\tilde\lambda_2)^{\dal}$. Expanding this in $z_{12}$, one finds the all-order collinear expansion
\be\label{gravcoll0}
M(1_+\, 2 \cdots  n) = -\sum_{p=0}^\infty z_{12}^{p-2}\sum_{q=0}^p\sum_{r=0}^q\frac{(-[1\p_2])^{q-r}}{r!(q-r)!}\sum_{i=3}^n\frac{[1i][1\p_i]^r}{z_{i2}^{p-1}}\,M(\mathbf{2}\cdots n)\,.
\ee
Naively, it appears that this expansion starts at order $1/z_{12}^2$ instead of the universal collinear pole $1/z_{12}$ expected in a (tree level) holomorphic collinear limit. 
This is actually a spurious pole that drops out due to momentum conservation.

To see this, recall that our $(n-1)$-point amplitude $M(\mathbf{2}\,3\cdots n)$ comes equipped with delta functions imposing the momentum conservation
\be
\mathbf{p}_2^{\al\dal} + p_3^{\al\dal} + \cdots + p_n^{\al\dal} = 0\,.
\ee
Contracting this with $\lambda_2^\al\tilde\lambda_1^{\dal}$ and using the fact that $\la i2\ra=z_{i2}$, we get the conservation law
\be
\sum_{i=3}^n\la i2\ra[1i]\equiv\sum_{i=3}^n z_{i2}[1i] = 0\,.
\ee
Hence, if we write out the term of order $1/z_{12}^2$ in the expansion \eqref{gravcoll0}, we find that it vanishes:
\be
-\frac{1}{z_{12}^2}\sum_{i=3}^nz_{i2}[1i]\,M(\mathbf{2}\cdots n) = 0\,.
\ee
This confirms that the collinear limit starts at order $1/z_{12}$.

So, dropping the $p=0$ term in the sum \eqref{gravcoll0}, we can shift the summation index $p\mapsto p+1$ to find the collinear expansion
\be\label{gravcoll}
\boxed{M(1_+\, 2 \cdots  n) = -\sum_{p=0}^\infty z_{12}^{p-1}\sum_{q=0}^{p+1}\sum_{r=0}^q\frac{(-[1\p_2])^{q-r}}{r!(q-r)!}\sum_{i=3}^n\frac{[1i][1\p_i]^r}{z_{i2}^{p}}\,M(\mathbf{2}\cdots n)\,.}
\ee
This is our result for the all-order collinear expansion of the tree level graviton MHV amplitude. Next let us turn this into OPE expansions.


\subsection{Celestial OPE of momentum eigenstates}

Like in gauge theory, it is expected that there exist local operators $G_{s_i}(z_i,\tilde\lambda_i)$ dual to graviton momentum eigenstates of helicity $s_i\in\{\pm2\}$ and spinor-helicity data $\lambda_i^\al=(1,z_i)$, $\tilde\lambda_i^{\dal}=\omega_i(1,\bz_i)$. Their correlators should compute graviton amplitudes in the bulk,
\be\label{duality1}
M(1\,2\cdots n) = \left\la G_{s_1}(z_1,\tilde\lambda_1)\,G_{s_2}(z_2,\tilde\lambda_2)\cdots G_{s_n}(z_n,\tilde\lambda_n)\right\ra\,.
\ee
Combining this with the collinear expansion \eqref{gravcoll}, we can once again extract the all-order OPE in the MHV sector. Let us begin by recalling the notion of soft graviton currents.

\paragraph{Soft graviton currents.} The soft expansion of a positive helicity graviton operator -- as valid at least within an MHV correlator -- is given by the double Taylor expansion\footnote{Again, this starts at order $\omega^0$ instead of the usual soft graviton pole $\omega^{-1}$ due to our choice of little group fixing.}
\be
G_+(z,\tilde\lambda) = \sum_{k,\ell=0}^\infty\frac{(\tilde\lambda^{\dot1})^k(\tilde\lambda^{\dot2})^\ell}{k!\,\ell!}\,\w[k,\ell](z)\,.
\ee
The chiral operators $\w[k,\ell](z)$ are currents with weights $h=2-\frac{k+\ell}{2}$ with respect to the conformal transformations of $z$. The currents at fixed $k+\ell$ also transform in a spin $\frac{k+\ell}{2}$ representation of the $\SL(2,\C)$ that rotates $\tilde\lambda^{\dal}$. As shown by Strominger \cite{Strominger:2021lvk}, the celestial OPEs of these currents form the loop algebra of Ham$(\C^2)$, the holomorphic symplectomorphisms of $\C^2$. In the celestial holography literature, Ham$(\C^2)$ is also referred to as the wedge subalgebra of $w_{1+\infty}$, which motivated the notation $\w[k,\ell]$ for the currents. The weight $h=2$ current $\w[0,0]$ actually turns out to be central, and its insertions in celestial correlators do not give rise to a nontrivial soft theorem. But it is usually left in for completeness.

The first step in deriving the celestial OPE is again to extract the singular terms in the OPE. The singular parts of the MHV collinear expansion are read off from \eqref{gravcoll},
\be
    M(1_+\,2\cdots n) = -\frac{1}{z_{12}}\left(\sum_{i=3}^n[1i] - \sum_{i=3}^n[1i][1\p_2]+\sum_{i=3}^n[1i][1\p_i]\right)M(\mathbf{2}\cdots n) + \rO(z_{12}^0)\,.
\ee
Since $\lambda_i^1=1$ in our choice of little group fixing, specific components of the $(n-1)$-point momentum and angular momentum conservation laws give the constraints
\begin{align}
    &\sum_{i=3}^n[1i]\,M(\mathbf{2}\cdots n) = -[12]\,M(\mathbf{2}\cdots n)\,,\\
    &\sum_{i=3}^n[1i][1\p_i]M(\mathbf{2}\cdots n) = -[12][1\p_2]M(\mathbf{2}\cdots n)\,,
\end{align}
thereby reducing the collinear limit to
\be
M(1_+\,2\cdots n) = \frac{[12]}{z_{12}}\,M(\mathbf{2}\cdots n) + \rO(z_{12}^0)\,.
\ee
Applying the duality dictionary \eqref{duality1}, the singular part of the hard graviton OPE is found to be
\be\label{gravope}
G_+(z_1,\tilde\lambda_1)\,G_{s_2}(z_2,\tilde\lambda_2) \sim \frac{[12]}{z_{12}}\,G_{s_2}(z_2,\tilde\lambda_1+\tilde\lambda_2)\,.
\ee
This is the graviton celestial OPE of \cite{Pate:2019lpp, Guevara:2021abz} written in momentum space. It is leading order in $z_{12}$ and all-order exact in $\bz_{12}$.

Setting $s_2=+2$, Taylor expanding both sides in $\tilde\lambda_1^{\dal}$ and $\tilde\lambda_2^{\dal}$, and equating coefficients of $(\tilde\lambda_1^{\dot1})^k(\tilde\lambda_1^{\dot2})^\ell(\tilde\lambda_2^{\dot1})^m(\tilde\lambda_2^{\dot2})^n$ yields the soft graviton current algebra (see Appendix~\ref{app:currentOPE} for details)
\be
\w[k,\ell](z)\,\w[m,n](w) \sim \frac{kn-\ell m}{z-w}\,\w[k+m-1,\ell+n-1](w)\,.
\ee
As mentioned above, this is the loop algebra of Ham$(\C^2)$.\footnote{Equip $\C^2$ with complex coordinates $v_1,v_2$ and a holomorphic Poisson structure
\begin{equation*}
\{f,g\} = \p_{v_1}f\,\p_{v_2}g-\p_{v_2}f\,\p_{v_1}g\,,\qquad f,g\in\Omega^0(\C^2)\,.    
\end{equation*}
Holomorphic symplectomorphisms of $\C^2$ are generated by the basis of holomorphic hamiltonians $v_1^kv_2^\ell$ with algebra $\{v_1^kv_2^\ell,v_1^mv_2^n\} = (kn-\ell m)\,v_1^{k+m-1}v_2^{\ell+n-1}$. The soft graviton currents span the loop algebra of this Poisson algebra.} It provides an infinite dimensional enhancement of the holographic symmetries of quantum gravity in flat space whose implications are still being fervently investigated. The origin of these symmetries is again found in the twistor theory of Penrose's nonlinear graviton construction \cite{Adamo:2021lrv}.

For our purposes, we will again be interested in working with their generating functions
\be
\w[r](z,\tilde\lambda) = \sum_{k=0}^r\frac{(\tilde\lambda^{\dot1})^k(\tilde\lambda^{\dot2})^{r-k}}{k!(r-k)!}\,\w[k,r-k](z)\,,\qquad r\in\Z_{\geq0}\,.
\ee
The hard gravitons can be expanded upon these as
\be
G_+(z,\tilde\lambda) = \sum_{r=0}^\infty\w[r](z,\tilde\lambda)\,,
\ee
where the $r^\text{th}$ term is essentially $\omega^r$ times the $r^\text{th}$ order soft graviton. From \eqref{gravope}, one finds a compact expression for the OPE of soft gravitons with hard gravitons,
\be\label{wG}
    \w[r](z_1,\tilde\lambda_1)\,G_{s_i}(z_i,\tilde\lambda_i) \sim \frac{[1i]}{z_{1i}}\,\frac{[1\p_i]^{r-1}}{(r-1)!}\,G_{s_i}(z_i,\tilde\lambda_i)\,.
\ee
The boost eigenbasis version of these soft-hard OPEs were found in \cite{Himwich:2021dau}. The central current $\w[0](z,\tilde\lambda)\equiv\w[0,0](z)$ of course has a regular OPE with all hard gravitons. The $r=1$ OPE is the supertranslation action. The $r=2$ OPE corresponds to superrotations. And the $r=3$ OPE corresponds to the sub-subleading soft graviton theorem of \cite{Cachazo:2014fwa}.

\paragraph{Soft graviton descendants.} Viewing $\w[r](z,\tilde\lambda)$ as chiral currents on the sphere with coordinate $z$, we define the soft graviton descendants of a hard graviton by the contour integral
\be
\w_{-p}[r](\tilde\lambda_1)G_{s_2}(z_2,\tilde\lambda_2) \vcentcolon= \oint_{|z_{12}|=\veps}\frac{\d z_1}{2\pi\im}\,\frac{1}{z_{12}^{p}}\,\w[r](z_1,\tilde\lambda_1)\,G_{s_2}(z_2,\tilde\lambda_2)
\ee
with $\veps\ll1$ and $p\geq0$. The $p=0$ term is just meant to be the primary occurring on the right of \eqref{wG}, while the $p\geq1$ descendants correspond to regular terms in the soft-hard OPE.

The current $\w[r](z,\tilde\lambda)$ has weight $h=2-r/2$ under conformal transformations of $z$, so correlators involving it can have a pole of order at most $r-4$ at $z=\infty$. Using \eqref{duality1}, \eqref{wG} and a contour-pulling argument that is identical to the gluon case, we then find that for $0\leq r\leq p+2$, an insertion of such a descendant in a correlator of hard gravitons can be expressed purely in terms of hard graviton amplitudes with one fewer particle:
\be
    \left\la \w_{-p}[r](\tilde\lambda_1)G_{s_2}(z_2,\tilde\lambda_2)\prod_{i=3}^n G_{s_i}(z_i,\tilde\lambda_i)\right\ra 
    = -\frac{1}{(r-1)!}\sum_{i=3}^n\frac{[1i][1\p_i]^{r-1}}{z_{i2}^p}\,M(2\,3\cdots n)\,.\label{descorrgrav}
\ee
After shifting $\tilde\lambda_2\mapsto\tilde\lambda_1+\tilde\lambda_2$ and $r\mapsto r+1$, these are immediately recognized to be the summands of the collinear expansion \eqref{gravcoll}.

\paragraph{All-order graviton OPE.} Using the duality dictionary \eqref{duality1} for graviton scattering, we can rewrite the MHV collinear expansion \eqref{gravcoll} as the OPE expansion
\begin{multline}
    \left\la G_+(z_1,\tilde\lambda_1)\,G_{s_2}(z_2,\tilde\lambda_2)\prod_{i=3}^n G_{s_i}(z_i,\tilde\lambda_i)\right\ra\\
    = \sum_{p=0}^\infty z_{12}^{p-1}\sum_{q=0}^{p+1}\sum_{r=0}^q\frac{(-[1\p_2])^{q-r}}{(q-r)!}\left\la \w_{-p}[r+1](\tilde\lambda_1)O^{a_2}_{s_2}(z_2,\tilde\lambda_1+\tilde\lambda_2)\prod_{i=3}^n O^{a_i}_{s_i}(z_i,\tilde\lambda_i)\right\ra\,.
\end{multline}
Again, $r+1\leq p+2$ so our contour manipulations work. This can be further simplified by exchanging the sums over $q$ and $r$, then shifting $r\mapsto r-1$. When the dust settles, one finds the following all-order OPE for two gravitons in the MHV sector:
\be\label{eq:gravitonopebox}
\boxed{G_+(z_1,\tilde\lambda_1)\,G_{s_2}(z_2,\tilde\lambda_2) = \sum_{p=0}^\infty z_{12}^{p-1}\sum_{r=1}^{p+2}\sum_{q=0}^{p+2-r}\frac{(-[1\p_2])^q}{q!}\,\w_{-p}[r](\tilde\lambda_1)G_{s_2}(z_2,\tilde\lambda_1+\tilde\lambda_2)\,.}
\ee
For the reader's benefit, we write out the first few terms of this OPE explicitly
\begingroup
\allowdisplaybreaks
\begin{align}
    &G_+(z_1,\tilde\lambda_1)\,G_{s_2}(z_2,\tilde\lambda_2) \nonumber\\
    =&  \frac{1}{z_{12}}\,\Bigl\{\bigl(1-[1\p_2])\,\w_0[1](\tilde\lambda_1)+\w_0[2](\tilde\lambda_1)\Bigr\}\,G_{s_2}(z_2,\tilde\lambda_1+\tilde\lambda_2)\nonumber\\
    &+ \biggl\{\bigg(1-[1\p_2]+\frac{[1\p_2]^2}{2!}\bigg)\,\w_{-1}[1](\tilde\lambda_1) + \bigl(1-[1\p_2]\bigr)\,\w_{-1}[2](\tilde\lambda_1) + \w_{-1}[3](\tilde\lambda_1)\biggr\}\,G_{s_2}(z_2,\tilde\lambda_1+\tilde\lambda_2)\nonumber\\
    &+ z_{12}\,\biggl\{\bigg(1-[1\p_2]+\frac{[1\p_2]^2}{2!}-\frac{[1\p_2]^3}{3!}\bigg)\,\w_{-2}[1](\tilde\lambda_1) + \bigg(1-[1\p_2]+\frac{[1\p_2]^2}{2!}\bigg)\,\w_{-2}[2](\tilde\lambda_1)\nonumber\\
    &\qquad\qquad\qquad+ \bigl(1-[1\p_2]\bigr)\,\w_{-2}[3](\tilde\lambda_1) + \w_{-2}[4](\tilde\lambda_1)\biggr\}\,G_{s_2}(z_2,\tilde\lambda_1+\tilde\lambda_2) + \rO(z_{12}^2)\,.
\end{align}
\endgroup
This helps illuminate the pattern of the sums. The first term looks different from \eqref{gravope}, but can be matched if one remembers from \eqref{wG} that the leading ``descendants'' are given by $\w_0[r](\tilde\lambda_1)G_{s_2}(z_2,\tilde\lambda_1+\tilde\lambda_2) = [12][1\p_2]^{r-1}G_{s_2}(z_2,\tilde\lambda_1+\tilde\lambda_2)/(r-1)!$. The remaining terms display the various orders of soft gravitons whose descendants enter the OPE. At the first subleading order, one finds supertranslation, superrotation as well as the sub-subleading soft graviton descendants. At order $z_{12}^{p-1}$, one only needs to use descendants of soft gravitons up to order $p+2$.

We find it remarkable that this can be expressed purely in terms of the soft graviton descendants (and $\bar L_{-1}=\p_{\bar z_2}$ descendants if expanded in small $\bz_{12}$). It does not really require an expansion in holomorphic conformal (or Virasoro) descendants. Moreover, explicit vertex algebra realizations of this OPE (like the ones for the gluon OPE coming from twistor strings \cite{Adamo:2022wjo}) have not been constructed as of the writing of this work. It would be very interesting to see if Skinner's gravitational twistor string \cite{Skinner:2013xp} could fill in this gap.


\subsection{Celestial OPE of boost eigenstates}

In a similar manner to that of Section~\ref{sec:gluonboost}, we can study the celestial OPE of graviton boost eigenstates via Mellin transforming the OPE of momentum eigenstates. The boost eigenstates of gravitons are defined in the same way as \eqref{eq:mellin}. Similar to \eqref{eq:softgluon}, we define the conformal primary positive helicity soft gravitons
\be\label{eq:softgraviton}
\w(z,\bz) \vcentcolon= \frac{\w(z,\tilde\lambda)}{(\veps\omega)^r} = \sum_{k=0}^r \frac{\bz^{r-k} \w[k,r-k](z)}{k!(r-k)!}\,.
\ee
One can then perform mode expansion over $z$ on $\w(z,\bz)$, whose OPE with the boost eigenstates can be obtained by contour integrals
\be
\w_{-p}[r](\bz) G^{\veps}_{\Delta,s}(w,\bar w) = \oint\frac{\d z}{2\pi\im}\,\frac{1}{(z-w)^p}\,\w[r](z,\bz)\,O^{\veps}_{\Delta,s}(w,\bar w) \,,
\ee
The modes $\w_{-p}[r](\bz)$ can be regarded as the generators of soft descendant states. Mellin transforming \eqref{eq:gravitonopebox} one can get the OPE of boost eigenstates of gravitons in terms of the following integral
\begin{equation}\label{eq:mellingraviton}
    \begin{aligned}
        & G^{\veps_1}_{\Delta_1,+}(z_1,\bz_1)\,G^{\veps_2}_{\Delta_2,s_2}(z_2,\bz_2) \\
        =& \sum_{p=0}^\infty z_{12}^{p-1}\sum_{r=0}^{p+1}\sum_{q=0}^{p+1-r}\sum_{m=0}^\infty\frac{(-1)^q}{q!m!}\int_0^\infty\d\omega_1\,\omega_1^{2\bh_1-1}\int_0^\infty\d\omega_2\,\omega_2^{2\bh_2-1} (\veps_1\omega_1)^{r+m+1}\\
        & \times \bigg(\frac{\veps_1\omega_1}{\veps_2\omega_2}\big(\bz_{12}\dbar_2+\omega_2\p_{\omega_2}\big)\bigg)^q\bz_{12}^m\dbar_2^m \w_{-p}[r+1](\bz_1) G^{\varepsilon_1}_{s_2}\bigg(z_2,\frac{(\veps_1\omega_1+\veps_2\omega_2)}{\veps_2\omega_2}\tilde\lambda_2\bigg).
    \end{aligned}
\end{equation}
After some math similar to Appendix~\ref{app:computeOPE}, the final OPE is
\begin{equation}\label{eq:gravitonOPE}
    \boxed{\begin{aligned}
        G^{\veps_1}_{\Delta_1,+}(z_1,\bz_1)\,G^{\veps_2}_{\Delta_2,s_2}(z_2,\bz_2) =& \sum_{p,\bar{m}=0}^\infty \sum_{r=0}^{p+1} \frac{\varepsilon_1^{r+1}}{\bar{m}!} \frac{B(2\hb_1+r+\bar{m}+1,2\hb_2)\,\Gamma(2\hb_1+p+3)}{(p-r+1)! \, \Gamma(2\hb_1+r+2)} \\
        & \times z_{12}^{p-1} \, \bz_{12}^{\bar{m}} \dbar_2^{\bar{m}} \, \w_{-p}[r+1](\bz_1) G^{\varepsilon_1}_{\Delta_1+\Delta_2+r-1,s_2}(z_2,\bz_2).
    \end{aligned}}
\end{equation}
The boost eigenstate OPE when $\varepsilon_1 = -\varepsilon_2$ is similar only that there are two terms produced corresponding to different ingoing/outgoing directions:
\begin{equation}\label{eq:gravitonOPEdifep}
    \boxed{\begin{aligned}
        & G^{\veps_1}_{\Delta_1,+}(z_1,\bz_1)\,G^{\veps_2}_{\Delta_2,s_2}(z_2,\bz_2) \\
        =& \sum_{p,\bar{m}=0}^\infty \sum_{r=0}^{p+1} \frac{\varepsilon_2^{r+1}}{\bar{m}!} (-1)^{r+\bar{m}+1} \frac{B(2\hb_1+r+\bar{m}+1,-2\hb_1-2\hb_2-r-\bar{m})\,\Gamma(2\hb_1+p+3)}{(p-r+1)! \, \Gamma(2\hb_1+r+2)} \\
        & \qquad\qquad\quad \times z_{12}^{p-1} \, \bz_{12}^{\bar{m}} \dbar_2^{\bar{m}} \, \w_{-p}[r+1](\bz_1) G^{\varepsilon_2}_{\Delta_1+\Delta_2+r-1,s_2}(z_2,\bz_2) \\
        &+ \sum_{p,\bar{m}=0}^\infty \sum_{r=0}^{p+1} \frac{\varepsilon_1^{r+1}}{\bar{m}!} \frac{B(-2\hb_1-2\hb_2-r-\bar{m},2\hb_2)\,\Gamma(2\hb_1+p+3)}{(p-r+1)! \, \Gamma(2\hb_1+r+2)} \\
        & \qquad\qquad\quad \times z_{12}^{p-1} \, \bz_{12}^{\bar{m}} \dbar_2^{\bar{m}} \, \w_{-p}[r+1](\bz_1) G^{\varepsilon_1}_{\Delta_1+\Delta_2+r-1,s_2}(z_2,\bz_2) \,.
    \end{aligned}}
\end{equation}

\section{Discussion}
\label{sec:disc}

Regular terms have proven useful for extracting differential equations satisfied by MHV amplitudes from null state decoupling relations in CCFT \cite{Banerjee:2020zlg,Banerjee:2020vnt}. Remarkably, these differential equations were already shown to be equivalent to a linearized version of tree level BCFW recursion in \cite{Hu:2021lrx}. Our calculation confirms that, indeed, the all-order celestial OPE is just a nonlinear extension of this idea that reproduces the exact recursion relation.

One intriguing shortcoming of our analysis is that the recursion relation does not allow us to obtain 3-point amplitudes from 2-point amplitudes. The latter are just meant to be the identity parts of the S-matrix, so that the corresponding 2-point celestial amplitudes tend to take the distributional form $\delta(z_{12})\delta(\bz_{12})$ \cite{Pasterski:2017ylz}. This means we do not yet know how to apply our OPE to a 2-point CCFT correlator and obtain the 3- and higher-point correlators. This is an important issue that needs to be resolved before any CCFT interpretation of 4d gauge theory or gravity may be taken seriously. A promising approach to this lies in the use of light and shadow transforms that may be cleverly used to make the 2- and 3-point functions non-distributional \cite{Crawley:2021ivb,Fan:2021isc,Sharma:2021gcz,Fan:2021pbp,Hu:2022syq,Banerjee:2022hgc,De:2022gjn,Chang:2022jut,Jorge-Diaz:2022dmy,Brown:2022miw}. It would be interesting to see if our all-order OPE can be used to build higher-point MHV amplitudes directly starting from a 2-point amplitude made non-distributional in this manner.

Another important issue facing celestial holography is to compute analogues of celestial OPE beyond the MHV sector. The singular part of the OPE is universal and valid for all scattering amplitudes, so it remains the same in N$^k$MHV amplitudes. At tree level, one expects that only the regular part may be nontrivially deformed.\footnote{Here, one includes terms that are regular in the holomorphic collinear limit but singular in the antiholomorphic one, and vice versa. It is standard to treat these limits as independent limits by complexifying momenta.} Our calculation of regular terms in the MHV OPE is meant to provide an accessible test case to develop methods for obtaining regular terms in more involved celestial OPEs. In this way, we hope that our results bring us one step closer to bootstrapping flat space amplitudes from constraints imposed by the existence of a celestial dual. Going beyond pure Yang-Mills, along the lines of e.g.~\cite{Luo:2014wea,Cachazo:2016njl}, or beyond tree level, along the lines of e.g.~\cite{Kalyanapuram:2020epb,Magnea:2021fvy}, would also be interesting.

At the first few orders, the gluon OPE was worked out in \cite{Ebert:2020nqf,Banerjee:2020vnt} and the graviton OPE in \cite{Banerjee:2020zlg}, where holomorphic conformal descendants were found. However, at those orders, due the existence of null states \cite{Banerjee:2020vnt,Banerjee:2020zlg,Banerjee:2021cly}, they can be fully converted to soft current descendants and antiholomorphic conformal descendants, consistent with our expressions (see for example \cite{Adamo:2022wjo}). Beyond the first few orders, consistency of our results with the appearance of holomorphic conformal descendants at arbitrarily high orders in $z$ and $\bar{z}$ \cite{Ebert:2020nqf} would require the existence of sufficiently many null states at each order. We leave this as an interesting question for the future. If the exact set of null states can be written down at each order, it would also allow one to write down the OPEs in alternative bases, with different forms useful for different situations.

Perhaps most ambitiously, if one found a way to reverse the logic of our paper and bootstrapped the all-order celestial OPE by other means, it would provide the means of bootstrapping BCFW recursion for MHV tree amplitudes directly from celestial CFT. Now, it turns out that MHV amplitudes as well as celestial OPEs have also been explored at loop level in \cite{Ball:2021tmb,Bittleston:2022jeq,Bhardwaj:2022anh} and in self-dual curved backgrounds in \cite{Adamo:2020syc,Costello:2022jpg,Bittleston:2023bzp}. These works discover that infinite dimensional symmetries like the $\w_{1+\infty}$ algebra persist in a wide range of scenarios, showing the robustness of the formalism of celestial holography. It may be possible to construct regular terms in the associated celestial OPEs, possibly using symmetry constraints, and derive novel recursion relations for (at least rational parts of) amplitudes at loop level or on curved backgrounds. Perhaps the best success story in this direction is the recent work by Costello on two-loop amplitudes in QCD-like theories \cite{Costello:2023vyy}, wherein we expect the choice of regular terms in the OPE to translate into a choice of conformal block.

\begin{acknowledgments}
We would like to thank Shamik Banerjee, Elizabeth Himwich, Yangrui Hu, Marcus Spradlin, Akshay Yelleshpur Srikant, and Anastasia Volovich for stimulating discussions. LR is supported by the US Department of Energy under contract DE-SC0010010 Task F. ASc is supported by the European Research Council (ERC) under the European Union’s Horizon 2020 research and innovation programme (grant agreement No 725110), \emph{Novel structures in scattering amplitudes}. ASh is supported by a Black Hole Initiative fellowship, funded by the Gordon and Betty Moore Foundation and the John Templeton Foundation. DW is supported by NSF grant PHY2107939.
\end{acknowledgments}

\begin{appendix}

\section{Poles at infinity}\label{app:sing}

In going from \eqref{descorr0} to \eqref{contourdef}, we deformed the contour and dropped the contribution from $z_1=\infty$. To justify this, we need to check whether the 1-form
\be\label{polarform}
\frac{\d z_1}{z_{12}^p}\,\left\la J^{a_1}[r](z_1,\tilde\lambda_1)\prod_{i=2}^n O^{a_i}_{s_i}(z_i,\tilde\lambda_i)\right\ra
\ee
has a pole at $z_1=\infty$ when viewed as a bulk scattering amplitude. We will show that this has no pole at $z_1=\infty$ precisely for the range
\be
0\leq r\leq p
\ee
that is suggested by the expected behavior of $J^{a_1}[r](z_1,\tilde\lambda_1)$ as a current of weight $h=1-r/2$. 

This is precisely what we will need for extracting the celestial OPE. We only show this explicitly in the gluon case. The corresponding proof in the graviton case involves identical steps taken using Hodges' formula \cite{Hodges:2012ym} for the MHV graviton amplitude (with rows and columns corresponding to graviton $1$ and the two negative helicity gravitons removed, and both reference spinors chosen to be $\lambda_1^\al$). We will avoid speculating what happens for $r>p$ (or $r>p+2$ in the graviton case), but this may be related to the residue of MHV amplitudes at infinity as obtained for instance in equation (35) of \cite{Guevara:2022qnm}.

To begin with the proof, we postulate as usual that there exist operators $O^{a_i}_{s_i}(z_i,\tilde\lambda_i)$ whose correlation function is the gluon MHV amplitude
\be
\left\la\prod_{i=1}^nO^{a_i}_{s_i}(z_i,\tilde\lambda_i)\right\ra = \sum_{\sigma\in S_{n-2}} C^{a_1a_2a_{\sigma_3}\dots a_{\sigma_n}}A[1\,2\,\sigma_3\cdots\sigma_n]\,,
\ee
where $A[1\,2\,3\cdots n]$ is the Parke-Taylor color-stripped amplitude
\be
A[1\,2\,3\cdots n] = \frac{z_{kl}^4}{z_{12}z_{23}\cdots z_{n1}}\,\delta^4(p_1+p_2+\cdots+p_n)
\ee
written for the configuration in which particles $k,l$ are negative helicity. For sake of simplicity, we are setting $\lambda_i^\al=(1,z_i),\tilde\lambda_i^{\dal}=\omega_i(1,\bz_i)$ from the start.

As before, take particle $1$ to be positive helicity without loss of generality. The soft-hard correlator
\be\label{softhard1}
\left\la J^{a_1}[r](z_1,\tilde\lambda_1)\prod_{i=2}^n O^{a_i}_{s_i}(z_i,\tilde\lambda_i)\right\ra
\ee
is obtained by expanding the MHV amplitude as a series in $\tilde\lambda_1^{\dal}$. At the level of the Parke-Taylor amplitude, one finds the series expansion
\be
A[1_+\,2\,3\cdots n] = \frac{z_{n2}}{z_{12}z_{n1}}\sum_{r=0}^\infty\frac{1}{r!}\,\cD^rA[2\,3\cdots n]\,,
\ee
where $\cD$ is the differential operator
\be\label{Ddef}
\cD = \frac{1}{z_{kl}}\left(z_{1l}[1\p_k]-z_{1k}[1\p_l]\right)\,.
\ee
This expansion may be obtained by using a Fourier representation of the momentum conserving delta functions,
\be
    \delta^4(p_1+p_2+\cdots+p_n) = \int\d^4x\sum_{r=0}^\infty\frac{(\im p_1\cdot x)^r}{r!}\,\e^{\im(p_2+\cdots+p_n)\cdot x}\,,
\ee
and applying the identity
\be
x^{\al\dal}\,\e^{\im(p_k+p_l)\cdot x} = \frac{-\im}{\la kl\ra}\left(\lambda_k^\al\frac{\p}{\p\tilde\lambda_{l\dal}}-\lambda_l^\al\frac{\p}{\p\tilde\lambda_{k\dal}}\right)\e^{\im(p_k+p_l)\cdot x}
\ee
to replace the factors of $(\im p_1\cdot x)^r$ by derivatives in the external data.

Since $\cD$ is homogeneous of degree 1 in $\tilde\lambda_1^{\dal}$ (and thereby also in $\omega_1$), \eqref{softhard1} gets contributions only from terms of order $\cD^r$ in each color order. This allows us to obtain the $r^\text{th}$ soft-limit
\be\label{softhard2}
\left\la J^{a_1}[r](z_1,\tilde\lambda_1)\prod_{i=2}^n O^{a_i}_{s_i}(z_i,\tilde\lambda_i)\right\ra = \sum_{\sigma\in S_{n-2}} C^{a_1a_2a_{\sigma_3}\dots a_{\sigma_n}}\frac{z_{\sigma_n2}}{z_{12}z_{\sigma_n1}}\,\frac{1}{r!}\,\cD^rA[2\,\sigma_3\cdots\sigma_n]
\ee
which has weight $r$ in $\omega_1$ as needed. (The actual soft limit is then obtained by dividing out the factor of $\omega_1^r$, but it is innocuous here so we leave it be.)

To complete the proof, we observe from \eqref{Ddef} that $\cD$ has a pole of order $1$ at $z_1=\infty$. So $\cD^r$ has a pole of order $r$. Because the extra factors of $z_{\sigma_n2}/z_{12}z_{\sigma_n1}$ have a second order zero, each summand in the net soft-hard correlator \eqref{softhard2} has a pole of order $r-2$ at $z_1=\infty$. On the other hand, the 1-form $\d z_1$ has a pole of order $2$ at infinity, while the prefactor $z_{12}^{-p}$ has a zero of order $p$. In total, \eqref{polarform} behaves like
\be
z_1^{r-2+2-p} = z_1^{r-p}
\ee
at large $z_1$. And as we wished to prove, this is regular at infinity for $r\leq p$. 

As an aside, notice that although the hard amplitude depended on $z_1$ in a distributional manner via its momentum conserving delta functions, the soft-hard amplitude only depends on $z_1$ rationally! So it makes sense to study its contour integrals as a function of $z_1$.

\section{Soft current OPEs}\label{app:currentOPE}

In this appendix, we obtain all-order OPEs between soft currents of both gluons and gravitons from momentum eigenstate hard-hard OPEs. Incidentally, the same results can be obtained from Mellin-transformed (boost eigenstate) OPEs.

\subsection{All-order soft gluon OPE}
Consider gluons first. From the definition \eqref{eq:soft_def}, the soft currents are given by its inverse form
\be
    J^a[m,n](z)=\oint\frac{\d\tilde{\lambda}^{\dot1}}{2\pi \im \tilde{\lambda}^{\dot1}}\frac{\d\tilde{\lambda}^{\dot2}}{2\pi \im \tilde{\lambda}^{\dot2}}(\tilde{\lambda}^{\dot1})^{-m}(\tilde{\lambda}^{\dot2})^{-n}\;m!\,n!\;O_+^a[z,\tilde\lambda].
\ee
Then the current OPE can be extracted by performing the integral
\begin{align}\label{eq:JJ_def}
    &J^a[k,l](z)J^b[m,n](w)\nonumber\\
    =&\oint\frac{\d\tilde{\lambda}^{\dot1}}{2\pi \im \tilde{\lambda}^{\dot1}}\frac{\d\tilde{\lambda}^{\dot2}}{2\pi \im \tilde{\lambda}^{\dot2}}
    \frac{\d\tilde{\kappa}^{\dot1}}{2\pi \im \tilde{\kappa}^{\dot1}}\frac{\d\tilde{\kappa}^{\dot2}}{2\pi \im \tilde{\kappa}^{\dot2}}
    \frac{k!\,l!\,m!\,n!}{
    (\tilde{\lambda}^{\dot1})^{k}(\tilde{\lambda}^{\dot2})^{l}(\tilde{\kappa}^{\dot1})^{m}(\tilde{\kappa}^{\dot2})^{n}}O_+^a[z,\tilde\lambda]O_+^b[w,\tilde\kappa].
\end{align}

Let us first look at the leading order calculation (in $z-w$), which already contains the main conceptual steps. The all-order calculation is analogous and will be presented afterwards. At leading order, the OPE between two positive-helicity hard gluons is given in \eqref{hardgluonope} with $s_2$ set to $+1$:
\be
O^{a}_+(z,\tilde\lambda)\,O^{b}_{+}(w,\tilde\kappa) \sim \frac{f^{ab}{}_c}{z-w}\;O^c_{+}(w,\tilde\lambda+\tilde\kappa).
\ee
The hard gluon $O^c_+(w,\tilde\lambda+\tilde\kappa)$ has the soft expansion
\begin{equation}
    O_+^c(w,\tilde\lambda+\tilde\kappa) = \sum_{u,v=0}^\infty\frac{(\tilde\lambda^{\dot1}+\tilde\kappa^{\dot1}){}^u(\tilde\lambda^{\dot2}+\tilde\kappa^{\dot2}){}^v}{u!\,v!}\;J^c[u,v](w)\,,
\end{equation}
which after binomial expansions becomes
\begin{equation}\label{eq:hard_binomial}
    O_+^c(w,\tilde\lambda+\tilde\kappa) =\sum_{u,v}^\infty \sum_{i=0}^u \sum_{j=0}^v {u \choose i}{v \choose j}(\tilde{\lambda}^{\dot1})^i(\tilde{\kappa}^{\dot1})^{u-i}(\tilde{\lambda}^{\dot2})^j(\tilde{\kappa}^{\dot2})^{v-j} \frac{J^c[u,v](w)}{u!v!}.
\end{equation}
This can now be substituted into \eqref{eq:JJ_def} to obtain
\begin{align}
    &J^a[k,l](z)J^b[m,n](w)\nonumber\\
    \sim&\,\frac{f^{ab}{}_c}{z-w}{k+m\choose k}{l+n\choose m}k!l!m!n!\frac{1}{(k+m)!(l+n)!}J^c[k+m,l+n](w)\nonumber\\
    =&\,\frac{f^{ab}{}_c}{z-w}J^c[k+m,l+n](w),
\end{align}
where we have used the fact that only the term with $i=k$, $u-i=m$, $j=l$ and $v-j=n$ gets picked up by the contour integrals.

Now repeat the steps while keeping to all orders. The all-order hard-hard gluon OPE was given in \eqref{eq:gluonopebox}. Since $J^a_{-p}[r](\tilde \lambda)$ appears in the OPE and depends on $\tilde\lambda$, we need to expand it out to perform the explicit contour integrations. Combining \eqref{eq:J_lamb_def} and \eqref{eq:Jp_def} leads to the expansion
\begin{equation}
    J^a_{-p}[r](\tilde \lambda)=\sum_{s=0}^r\frac{(\tilde\lambda^{\dot1})^s(\tilde\lambda^{\dot2})^{r-s}}{s!\,(r-s)!}\;J^a_{-p}[s,r-s].
\end{equation}
Another object in \eqref{eq:gluonopebox} that needs expansion is
\begin{equation}\label{eq:diff_binomial}
    [\tilde \lambda\partial_{\tilde\kappa}]^q=\sum_{t=0}^q{q\choose t}\left(\tilde\lambda^{\dot 1}\frac{\partial}{\partial\tilde\kappa^{\dot 1}}\right)^t\left(\tilde\lambda^{\dot 2}\frac{\partial}{\partial\tilde\kappa^{\dot 2}}\right)^{q-t}.
\end{equation}
With that, the soft current OPE can now be easily obtained. Substituting \eqref{eq:Jp_def}, \eqref{eq:diff_binomial} and \eqref{eq:hard_binomial} into \eqref{eq:gluonopebox}, and then the resulting expression into \eqref{eq:JJ_def}, we obtain
\begin{align}\label{eq:gluoncurrentOPElong}
    &J^a[k,l](z)J^b[m,n](w)\nonumber\\
    =&\sum_{p=0}^\infty (z-w)^{p-1}\sum_{r=0}^p\sum_{q=0}^{p-r}\sum_{s=0}^r\sum_{t=0}^q(-1)^q{k\choose s}{l\choose r-s}{k-s\choose t}{l-(r-s)\choose q-t}
    \nonumber\\
    &\times J^a_{-p}[s,r-s]J^b[k+m-s,l+n-(r-s)](w).
\end{align}
Similar to before, to arrive at this expression, we have performed the contour integrals which picked up terms satisfying
\begin{equation}
    s+i+t=k, \quad u-i-t=m,\quad
    r-s+j+q-t=l,\quad v-j-q+t=n.
\end{equation}
Not all terms in the sums necessarily contribute. In fact, during the derivation, one has to be careful with two types of conditions. One arises from the fact that taking too many derivatives (with respect to $\tilde\kappa^{\dot 1}$ and $\tilde\kappa^{\dot 2}$ in our case) would result in zero, and this requires $t\le u-i$ and $q-t\le v-j$. The other type of conditions comes from the ranges of $u$, $v$, $i$ and $j$ in \eqref{eq:hard_binomial}. After carefully treating them, we find that only terms with
\begin{equation}
    s\le k+m,\quad r-s\le l+n,\quad s+t\le k,\quad r-s+q-t\le l
\end{equation}
contribute to the sum. The formula written above is, however, still correct, owning to the fact that $J^a[m,n]=0$ when either $m$ or $n$ is negative and that ${m\choose n}=0$ when $m<n$. 

Performing the summations over $q$ and $t$ leads to
\begin{equation}
\boxed{
\begin{aligned}
    J^a[k,l](z)J^b[m,n](w)
    =&\sum_{p=0}^\infty (z-w)^{p-1}\sum_{r=0}^p\sum_{s=0}^r
    {k\choose s}{l\choose r-s}{p-k-l\choose p-r}
    \\
    &\times
     J^a_{-p}[s,r-s]J^b[k+m-s,l+n-(r-s)](w).
\end{aligned}
}
\end{equation}
If desired, this can be written in an alternative form by swapping the order of summations and relabeling, leading to
\begin{align}
    J^a[k,l](z)J^b[m,n](w)
    =&\sum_{p=0}^\infty (z-w)^{p-1}\sum_{s=0}^p\sum_{r=0}^{p-s}
    {k\choose s}{l\choose r}{p-k-l\choose p-r-s}
    \nonumber\\
    &\times
     J^a_{-p}[s,r]J^b[k+m-s,l+n-r](w).
\end{align}

\subsection{All-order soft graviton OPE}
The soft graviton current OPE can be derived in exactly the same way, with the analogue of \eqref{eq:gluoncurrentOPElong} being
\begin{align}
    &\w[k,l](z)\w[m,n](w)\nonumber\\
    =&\sum_{p=0}^\infty (z-w)^{p-1}\sum_{r=1}^{p+2}\sum_{q=0}^{p+2-r}\sum_{s=0}^r\sum_{t=0}^q(-1)^q{k\choose s}{l\choose r-s}{k-s\choose t}{l-(r-s)\choose q-t}
    \nonumber\\
    &\times \w_{-p}[s,r-s]\w[k+m-s,l+n-(r-s)](w).
\end{align}
Performing the summations over $q$ and $t$ gives
\begin{equation}
\boxed{
\begin{aligned}
    \w[k,l](z)\w[m,n](w)
    =&\sum_{p=0}^\infty (z-w)^{p-1}\sum_{r=1}^{p+2}\sum_{s=0}^r
    {k\choose s}{l\choose r-s}{p+2-k-l\choose p+2-r}
    \\
    &\times\w_{-p}[s,r-s]\w[k+m-s,l+n-(r-s)](w).
\end{aligned}
}
\end{equation}
Unlike the gluon case, the $r$ and $s$ sums cannot be swapped to give a simpler expression. 

\section{Mellin transforms of momentum space OPEs}\label{app:computeOPE}

In this section we provide the computation details for the OPE of boost eigenstates when $\varepsilon_1 = \varepsilon_2$. Starting from \eqref{eq:mellingluon}, we first use the identity:
\begin{align}\label{eq:divideq}
        & \bigg(\frac{\veps_1\omega_1}{\veps_2\omega_2}\big(\bz_{12}\dbar_2+\omega_2\p_{\omega_2}\big)\bigg)^q \nonumber\\
        =& \sum_{q'=0}^{q} \frac{q!}{q'!\, (q-q')!} \left( \frac{\varepsilon_1 \omega_1}{\varepsilon_2 \omega_2} \omega_2 \pa_{\omega_2} \right)^{q-q'} \left( \frac{\varepsilon_1 \omega_1}{\varepsilon_2 \omega_2} \right)^{q'} \bz_{12}^{q'} \dbar_2^{q'} \nonumber\\
        =& \sum_{q'=0}^{q} \frac{q!}{q'!\, (q-q')!} \left( \frac{\varepsilon_1 \omega_1}{\varepsilon_2 \omega_2} \right)^q \Big( \omega_2\pa_{\omega_2} - q + 1 \Big) \cdots \Big( \omega_2\pa_{\omega_2} - q' \Big)\, \bz_{12}^{q'} \dbar_2^{q'} \,.
\end{align}
Now we can see there are two contributions of $\dbar_2$: one is from $\bz_{12}^{q'}\dbar_2^{q'}$ above, the other is from $\bz_{12}^{m} \dbar_2^{m}$ in \eqref{eq:mellingluon}. We prefer to merge them together:
    \begin{align}\label{eq:mergebz}
        \big( \bz_{12}^{q'} \dbar_2^{q'} \big) \big( \bz_{12}^{m}\dbar_2^{m} \big)
        &= \sum_{s=0}^{q'} \binom{q'}{s} (-1)^{s} m \cdots (m-s+1) \bz_{12}^{q'}\, \bz_{12}^{m-s}\, \dbar_2^{q'-s}\, \dbar_2^{m}\nonumber \\
        &= \sum_{s=0}^{q'} (-1)^{s} \frac{q'!\, m!}{s! (q'-s)! (m-s)!} \bz_{12}^{m-s} \dbar_2^{q_2-s} \,.
    \end{align}
Plugging \eqref{eq:divideq} and \eqref{eq:mergebz} into \eqref{eq:mellingluon}, we get
\begingroup
\allowdisplaybreaks
\begin{align}\label{eq:gluonOPEmid}
    & O^{\veps_1,a_1}_{\Delta_1,+}(z_1,\bz_1)\,O^{\veps_2,a_2}_{\Delta_2,s_2}(z_2,\bz_2) \nonumber\\
    =& \sum_{p=0}^\infty z_{12}^{p-1}\sum_{r=0}^p\sum_{q=0}^{p-r}\sum_{m=0}^\infty \sum_{q'=0}^{q} \sum_{s=0}^{q'} \int_0^\infty\d\omega_1\,\omega_1^{2\bh_1-1}\int_0^\infty\d\omega_2\,\omega_2^{2\bh_2-1}\nonumber\\
    & \times \frac{(-1)^{q-s}}{q'!\, s!(q'-s)! (m-s)!} \frac{(\veps_1\omega_1)^{q+r} }{(\varepsilon_2 \omega_2)^{q}} \left(\omega_2\pa_{\omega_2} + 1 \right) \cdots \left(\omega_2\pa_{\omega_2} - q' \right)\nonumber\\
    & \times \left( \frac{\varepsilon_1 \omega_1}{\varepsilon_1 \omega_1 + \varepsilon_2 \omega_2} \right)^m \bz_{12}^{q'+m-s} \dbar_2^{q'+m-s} J^{a_1}_{-p}[r](\bz_1)O^{a_2}_{s_2}\bigg(z_2,\frac{(\veps_1\omega_1+\veps_2\omega_2)}{\veps_2\omega_2}\tilde\lambda_2\bigg) \nonumber\\
    =& \sum_{p=0}^\infty z_{12}^{p-1}\sum_{r=0}^p\sum_{q=0}^{p-r}\sum_{m=0}^\infty \sum_{q'=0}^{q} \sum_{s=0}^{q'} \int_0^\infty\d\omega_1\,\omega_1^{2\bh_1-1}\int_0^\infty\d\omega_2\,\omega_2^{2\bh_2-1}\nonumber\\
    & \times \frac{(-1)^{q-s}}{q'!\, s!(q'-s)! (m-s)!} \frac{(\veps_1\omega_1)^{q+r} }{(\varepsilon_2 \omega_2)^{q}} \left(-2\hb_2 + 1 \right) \cdots \left(-2\hb_2 + q' \right)\nonumber\\
    & \times \left( \frac{\varepsilon_1 \omega_1}{\varepsilon_1 \omega_1 + \varepsilon_2 \omega_2} \right)^m \bz_{12}^{q'+m-s} \dbar_2^{q'+m-s} J^{a_1}_{-p}[r](\bz_1)O^{a_2}_{s_2}\bigg(z_2,\frac{(\veps_1\omega_1+\veps_2\omega_2)}{\veps_2\omega_2}\tilde\lambda_2\bigg) \,.
\end{align}
\endgroup
Taking $\varepsilon_1 = \varepsilon_2$ and use the change of integration variables
\be
\omega_1 = t\omega\,,\quad\omega_2=(1-t)\omega\,,\quad t\in(0,1)\,,\quad\omega\in(0,\infty)\,,
\ee
the OPE \eqref{eq:gluonOPEmid} turns into:
\begin{equation}
    \begin{aligned}
        & O^{\veps_1,a_1}_{\Delta_1,+}(z_1,\bz_1)\,O^{\veps_2,a_2}_{\Delta_2,s_2}(z_2,\bz_2) \\
        =& \sum_{p=0}^\infty z_{12}^{p-1}\sum_{r=0}^p\sum_{q=0}^{p-r}\sum_{m=0}^\infty \sum_{q'=0}^{q} \sum_{s=0}^{q'} \int_{0}^{\infty} \d\omega\ \omega^{2\hb_1+2\hb_2-1+r} \int_{0}^{1} \d t \ t^{2\hb_1-1+r+q+m} \, (1-t)^{2\hb_2-1-q} \\
        &\times \frac{\varepsilon_1^{r}\, (-1)^{q-s} (-2\hb_2+q-q')!}{(q-q')! s!\, (q'-s)! (m-s)! (-2\hb_2)!} \bz_{12}^{q'+m-s} \dbar_2^{q'+m-s} J^{a_1}_{-p}[r](\bz_1)O^{a_2}_{s_2}\bigg(z_2,\frac{(\veps_1\omega_1+\veps_2\omega_2)}{\veps_2\omega_2}\tilde\lambda_2\bigg), \\
    \end{aligned}
\end{equation}
which equals \eqref{eq:gluonOPEmid2}. The $t$-integral produces an Euler-Beta function:
\begin{equation}
    \begin{aligned}
        & O^{\veps_1,a_1}_{\Delta_1,+}(z_1,\bz_1)\,O^{\veps_2,a_2}_{\Delta_2,s_2}(z_2,\bz_2) \\
        =& \sum_{p=0}^\infty z_{12}^{p-1}\sum_{r=0}^p\sum_{q=0}^{p-r}\sum_{m=0}^\infty \sum_{q'=0}^{q} \sum_{s=0}^{q'} \int_{0}^{\infty} \d\omega\ \omega^{2\hb_1+2\hb_2-1+r} \, \frac{\Gamma(2\hb_2 - q) \Gamma(2\hb_1+r+q+m)}{\Gamma(2\hb_1 + 2\hb_2 + r+m)}\\
        &\times \frac{\varepsilon_1^{r}\, (-1)^{q-s} (-2\hb_2+q-q')!}{(q-q')! s!\, (q'-s)! (m-s)! (-2\hb_2)!} \bz_{12}^{q'+m-s} \dbar_2^{q'+m-s} J^{a_1}_{-p}[r](\bz_1)O^{a_2}_{s_2}\bigg(z_2,\frac{(\veps_1\omega_1+\veps_2\omega_2)}{\veps_2\omega_2}\tilde\lambda_2\bigg). \\
    \end{aligned}
\end{equation}
Since the total power of $\bz_{12}$ and $\dbar_2$ in the summand is equal to $q'+m-s$, we replace the index $m$ by $\bar{m} := q'+m-s$. Furthermore, we also consider changing $q \to b := q-q'$, $q' \to a:=q'-s$. Therefore the regions of summation above become
\begin{align}\label{eq:gluonOPE2}
        & O^{\veps_1,a_1}_{\Delta_1,+}(z_1,\bz_1)\,O^{\veps_2,a_2}_{\Delta_2,s_2}(z_2,\bz_2) \nonumber\\
        =& \sum_{p=0}^\infty z_{12}^{p-1}\sum_{r=0}^p \sum_{\bar{m}=0}^{\infty} \sum_{a=0}^{\text{min}(\bar{m},p-r)} \sum_{s=0}^{p-r-a}\, \sum_{b=0}^{p-r-a-s} \int_{0}^{\infty} \d\omega\ \omega^{2\hb_1+2\hb_2-1+r} \nonumber\\
        & \times (-1)^{a+b} \varepsilon_1^{r}\, \frac{\Gamma(2\hb_2-a-b-s) \Gamma(2\hb_1+r+b+s+\bar{m}) \, \Gamma(-2\hb_2+b+1)}{\Gamma(2\hb_1+2\hb_2+r+\bar{m}-a)\, b!\, s!\, a!\, (\bar{m}-a-s)! \, \Gamma(-2\hb_2+1)} \nonumber\\
        &  \times\bz_{12}^{\bar{m}} \dbar_2^{\bar{m}} J^{a_1}_{-p}[r](\bz_1)O^{a_2}_{s_2}\bigg(z_2,\frac{(\veps_1\omega_1+\veps_2\omega_2)}{\veps_2\omega_2}\tilde\lambda_2\bigg) \,.
    \end{align}
To perform the summation on $a,b,s$, we use the identity
\begin{align}\label{eq:OPEid}
        & \sum_{a=0}^{\text{min}(\bar{m},p-r)} \sum_{s=0}^{p-r-a}\, \sum_{b=0}^{p-r-a-s} (-1)^{a+b} \frac{\Gamma(2\hb_2-a-b-s) \Gamma(2\hb_1+r+b+s+\bar{m})\, \Gamma(-2\hb_2+b+1)}{\Gamma(2\hb_1+2\hb_2+r+\bar{m}-a)\, b!\, s!\, a!\, (\bar{m}-a-s)! \, \Gamma(-2\hb_2+1)} \nonumber\\
        &=\, \frac{1}{\bar{m}!} B(2\hb_1+r+\bar{m}, 2\hb_2) \frac{\Gamma(2\hb_1+p+1)}{(p-r)!\, \Gamma(2\hb_1+r+1)}\,.
\end{align}
Plugging \eqref{eq:OPEid} into \eqref{eq:gluonOPE2}, we obtain the final answer \eqref{eq:gluonOPE}. In a similar manner, we can also obtain the Mellin-transformed graviton OPE \eqref{eq:gravitonOPE}.

\end{appendix}

\bibliographystyle{JHEP}
\bibliography{library}

\end{document}